\documentclass[twocolumn,twocolappendix]{aastex631}
\usepackage{hyperref,graphicx}
\usepackage{scalerel,stackengine}
\usepackage{enumitem}
\usepackage{amsmath,amssymb}

\usepackage{subcaption}
\usepackage{natbib}

\begin{document}
\shorttitle{Metric Polarization Calibration using Unpolarized Sky}
\shortauthors{Kansabanik et al.}

\title{A Formalism for Calibrating the Instrumental Polarization of Radio Interferometric Arrays at Meter Wavelengths using Unpolarized Sky: A Demonstration using the MWA Observations}
\correspondingauthor{Devojyoti Kansabanik}
\email{dkansabanik@ucar.edu, devojyoti96@gmail.com}

\author[0000-0001-8801-9635]{Devojyoti Kansabanik}
\affiliation{Cooperative Programs for the Advancement of Earth System Science, University Corporation for Atmospheric Research, 3090 Center Green Dr, Boulder, CO, USA 80301}
\affiliation{The Johns Hopkins University Applied Physics Laboratory, 11001 Johns Hopkins Rd, Laurel, USA 20723}
\author[0000-0002-8164-5948]{Angelos Vourlidas}
\affiliation{The Johns Hopkins University Applied Physics Laboratory, 11001 Johns Hopkins Rd, Laurel, USA 20723}
\author[0009-0006-3517-2031]{Soham Dey}
\affiliation{National Centre for Radio Astrophysics, Tata Institute of Fundamental Research, S. P. Pune University Campus, Pune 411007, India}
\author[0000-0002-2325-5298]{Surajit Mondal}
\affiliation{Center for Solar-Terrestrial Research, New Jersey Institute of Technology, 323 M L King Jr Boulevard, Newark, NJ 07102-1982, USA}
\author[0000-0002-4768-9058]{Divya Oberoi}
\affiliation{National Centre for Radio Astrophysics, Tata Institute of Fundamental Research, S. P. Pune University Campus, Pune 411007, India}

\begin{abstract}
Calibration of instrumental polarization is critical for measuring polarized radio emissions from astrophysical sources to extract the magnetic field information in astrophysical, heliospheric, and terrestrial plasmas. At meter wavelengths, calibration of radio polarimetric observations is particularly challenging because of the scarcity of bright polarized sources due to significant Faraday depolarization. Here, we present a novel formalism for polarization calibration using an unpolarized sky model. The formalism is specifically designed for wide-field, low-frequency instruments like the Murchison Widefield Array (MWA), the LOw Frequency ARray (LOFAR), New Extension in Nançay Upgrading LoFAR (NenuFAR), Owens Valley Radio Observatory - Long Wavelength Array (OVRO-LWA), low-frequency telescope of the Square Kilometre Array Observatory (SKAO-low), etc. By leveraging the apparent polarization of the unpolarized sky induced by the polarized primary beam of the radio telescope, this method avoids dependence on bright polarized calibrators. It is also immune to ionospheric Faraday rotation. The validation of the approach via MWA observations confirms the accuracy of the method. This formalism provides a robust framework for low-frequency polarization calibration. It addresses the longstanding calibration challenges and advances the field of low-frequency polarimetry by enabling polarization studies of astrophysical radio sources.
\end{abstract}

\section{Introduction}\label{sec:intro}
The polarization of electromagnetic radiation provides crucial insights into the properties of the radio sources emitting that radiation and the intervening medium between the source and observer across all wavelengths. At radio wavelengths, polarization carries the imprint of the magnetic field of the source and/or the intervening medium. Hence, radio polarimetry is an important observational tool for probing magnetic fields of terrestrial to astrophysical plasmas ranging from the terrestrial ionosphere, solar corona, and heliosphere, other stars and exoplanets, interstellar medium, galaxies, intergalactic medium and clusters of galaxies. Radio telescopes typically have two orthogonally polarized antennas sensitive to measuring polarization of incident radio emission. This design allows most of the radio telescopes in the world to routinely conduct polarimetric observations. However, accurate utilization of these polarization measurements for probing terrestrial and astrophysical plasmas necessitates careful correction for instrumental polarization.

The polarization state of electromagnetic radiation can be represented as a vector using Stokes parameters $I,\ Q,\ U,\ V$ \citep{stokes1851}. Stokes $I$ represents the total intensity of the electromagnetic radiation, and the polarization state of the emission is represented by the polarization vector, $p=(Q,\ U,\ V)$ in a {\it pol-vector} space ($\hat{Q}, \hat{U}, \hat{V}$). Instrumental polarization can alter both the length and the direction of $p$. Errors in the length of $p$ arise from mixing with Stokes $I$, a phenomenon known as {\it pol-conversion} \citep{Hamaker2000}. On the other hand, any mixing between Stokes $Q, U, V$ results in a 3-dimensional rotation of $p$ in the polarization vector space, keeping its length constant -- this is referred to as {\it pol-rotation} \citep{Hamaker2000}. To uniquely calibrate the length and direction of $p$ as measured by a radio interferometer, three reference source models with linearly independent Stokes parameters in the {\it pol-vector} space are required \citep{Hamaker1996_1, Sault1996_2}.

In the centimeter wavelength and microwave regimes, numerous bright polarized astronomical sources are available to serve as polarization calibrators. 
To calibrate the absolute polarization angle of these sources, various planetary bodies such as the Moon, Mars, and Venus have been used \citep{Perley_2013}, since their polarization angle properties are well understood from fundamental physics \citep{Heiles1963}. However, at meter-wavelengths, the depolarization caused by Faraday rotation (FR) at the source or through the intervening medium reduces polarized flux density as well as complicates the spectral structure of the source model. 
Additionally, planetary bodies exhibit very low or nearly zero flux densities at meter wavelengths, rendering them unsuitable for absolute polarization calibration. These factors contribute to the challenges of absolute polarization calibration at meter wavelengths \citep{lenc2017}.

Scientifically, low-frequency polarimetry provides insights into key questions in astronomy, including low-frequency polarized source counts, depolarization effects, emission mechanisms of pulsars, magnetars, fast radio bursts (FRBs), stellar flares, space weather, exoplanet detection, and magnetic fields in the Milky Way, interstellar medium, and solar system. Despite the calibration challenges, low-frequency polarimetry has been successfully demonstrated \citep{lenc2017, Risley2018, Risley2020} using the Murchison Widefield Array \citep[MWA,][]{Tingay2013,Wayth2018} and LOw Frequency ARray \citep[LOFAR,][]{lofar2013}. These studies relied on bright linearly polarized sources and the stability of instrumental properties, which provided adequate accuracy for surveys in which small temporal changes of instrumental effects are averaged out over multiple epochs. However, this approach is insufficient for bright sources like the Sun, where even minimal residual instrumental polarization leads to detectable systematics, and for transient sources such as FRBs or solar coronal mass ejections (CMEs), where polarization calibration over shorter time intervals is essential.

A novel method for calibrating instrumental polarization at meter wavelengths using an unpolarized sky model for wide field-of-view (FoV) instruments was initially proposed by \cite{Farnes2014} for the legacy Giant Metrewave Radio Telescope \citep[GMRT,][]{Swarup1991}, and more recently applied by \cite{Byrne2022} (hereafter BY22) for the MWA. However, these studies did not establish the general applicability of the method nor did they assess its accuracy in calibrating instrumental polarization. Consequently, polarization calibration at meter wavelengths still relies on bright polarized sources \citep{lenc2017}. This paper aims to establish the applicability of this method through a coherent mathematical framework known as the ``Radio Interferometric Measurement Equation" (RIME) \citep{Hamaker1996_1, Hamaker2000}, and to demonstrate its efficacy and accuracy using several MWA observations. 

The paper is organized as follows. Section \ref{sec:theory} provides a brief overview of the relevant mathematical background of RIME. The formalism of the proposed method is described in Section \ref{sec:formalism}. In Section \ref{sec:key_conditions}, we outline the essential conditions that must be met for this formalism to be applicable. Section \ref{sec:MWA_calibration} presents the steps for calibrating the MWA observations followed by results and demonstrates the accuracy using the MWA observations in Section \ref{sec:mwa_obs_results}. We discuss the significance and general applicability of this method in Section \ref{sec:discussion}, followed by conclusions in Section \ref{sec:conclusion}.

\section{A Brief Overview of the  Mathematical Framework of RIME}\label{sec:theory}
Radio interferometers comprise several antennas. Each of them measures voltages, $\vec{v}$ = ($v_\mathrm{a},\ v_\mathrm{b}$) corresponding to the two orthogonally polarized components of the incident radio electromagnetic field, $\vec{E}$ = ($E_\mathrm{a},\ E_\mathrm{b}$). The measured voltages can be expressed in terms of incident electric field, $\vec{E}$ and antenna-based Jones matrix, $J$, \citep{Jones1941} as:
\begin{equation}\label{eq:jones_matrix}
\begin{split}
    \vec{v} &=J \ \Vec{E}\\
    \begin{pmatrix}\mathcal{V}_\mathrm{a}\\\mathcal{V}_\mathrm{b}\end{pmatrix} &=J \begin{pmatrix}E_\mathrm{a}\\E_\mathrm{b}\end{pmatrix},
\end{split}
\end{equation}
where, $a,\ b$ represent two mutually orthogonal polarization and $J$ is a 2$\times$2 matrix representing instrumental and atmospheric effects. The fundamental observables for a radio interferometer are the correlation products of $\vec{v}$ between two antennas, known as {\it visibilities}. Observed visibilities between two antennas ($j,\ k$) and two polarizations ($a,\ b$) can be expressed as a 2$\times$2 {\it Visibility Matrix}:
\begin{equation}\label{eq:visibility_matrix}
    \begin{split}
      \mathcal{V}_\mathrm{jk}^\prime&= 2\begin{pmatrix} \mathcal{V}_\mathrm{aj}\mathcal{V}_\mathrm{ak}^* & \mathcal{V}_\mathrm{aj}\mathcal{V}_\mathrm{bk}^*\\\mathcal{V}_\mathrm{bj}\mathcal{V}_\mathrm{ak}^* & \mathcal{V}_\mathrm{bj}\mathcal{V}_\mathrm{bk}^*
        \end{pmatrix}\\
        &=2J_\mathrm{j}
        \begin{pmatrix}
        E_\mathrm{aj}E_\mathrm{ak}^* & E_\mathrm{aj}E_\mathrm{bk}^*\\E_\mathrm{bj}E_\mathrm{ak}^* & E_\mathrm{bj}E_\mathrm{bk}^*
        \end{pmatrix}
       J_\mathrm{k}^\dagger\\
       \mathcal{V}_\mathrm{jk}^\prime&=J_\mathrm{j}\begin{pmatrix}
       \mathcal{V}_I+\mathcal{V}_Q & \mathcal{V}_U+i\mathcal{V}_V \\ \mathcal{V}_U-i\mathcal{V}_V & \mathcal{V}_I-\mathcal{V}_Q
       \end{pmatrix}_\mathrm{jk}J_\mathrm{k}^\dagger= J_\mathrm{j}\ \mathcal{V}_\mathrm{jk}\ J_\mathrm{k}^\dagger,
    \end{split}
\end{equation}
where, $\mathcal{V}_\mathrm{jk}$ represents the true source visibility matrix, while $\mathcal{V}_\mathrm{jk}^\prime$ is the observed visibility matrix, and $\mathcal{V}_I,\ \mathcal{V}_Q,\ \mathcal{V}_U,\ \mathcal{V}_V$ are the Stokes visibilities of the incident radiation following the IAU/IEEE Stokes parameter definition \citep{IAU_1973, Hamaker1996_3}. Equation \ref{eq:visibility_matrix} is known as the RIME. The goal of polarization calibration for a radio interferometer is to estimate the Jones matrices ($J_\mathrm{j}$) for each antenna. This is usually done using a known sky model, $\mathcal{V}_\mathrm{jk}$, and the observed $\mathcal{V}_\mathrm{jk}^\prime$. Although the Jones matrices consist of eight real numbers (four complex values), there are effectively 7 degrees of freedom (def)\citep{Sault1996_2}. Calibration using a source with a known Stokes vector (whether polarized or unpolarized) constrains four of these parameters, but the remaining three parameters, which relate to {\it pol-rotation}, are the most challenging to determine. This is described in detail in Section \ref{subsec:jip_dter}. These unconstrained parameters result in the rotation of the polarization vector $p$ within the polarization vector space \citep{Hamaker2000}.

Analogously to the Euler angle representation of 3-dimensional rotation in classical mechanics, the 3-dimensional rotation in polarization vector space can be described using another Euler angle representation involving rotations along the three Stokes axes ($Q,\ U,\ V$) using three linearly independent 2$\times$2 unitary Jones matrices. The specific parametric forms of these Jones matrices depend on the instrumental polarization basis (linear or circular) \citep{Hamaker2000}. To constrain these three Jones matrices, at least three sources with linearly independent Stokes vectors are required. This has been achieved at GHz frequencies \citep{evla_memo_201_hales2017} using an unpolarized source along with two linearly polarized sources with different position angles. Another approach involves observing a single linearly polarized source over a wide range of parallactic angles, where the rotation of the instrumental beam with respect to the sky coordinates causes apparent changes in the source polarization, effectively providing linearly independent Stokes vectors to constrain the Jones matrices.

\begin{figure}[!htbp]
    \centering
    \includegraphics[width=0.47\textwidth]{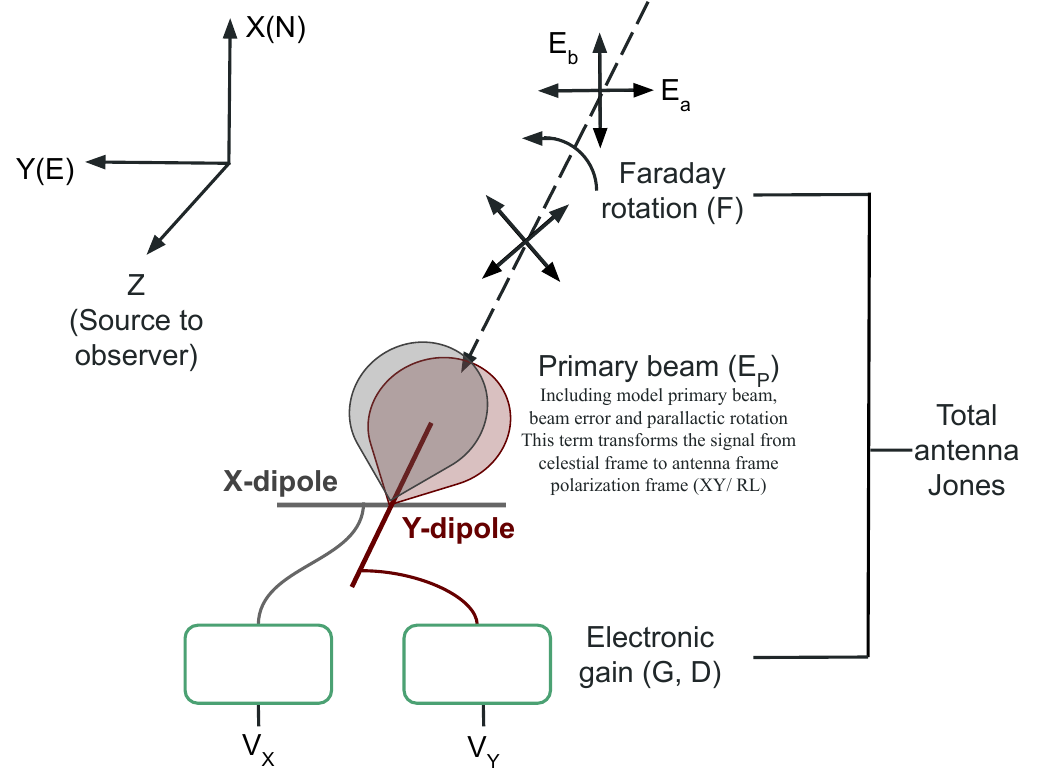}
    \caption{A schematic diagram illustrating the propagation of electromagnetic radiation as it undergoes various atmospheric and instrumental effects, along with the associated Jones terms described in Equation \ref{eq2}.}
    \label{fig:jones_chain}
\end{figure}

\section{Proposed Formalism}\label{sec:formalism}
The electromagnetic signal from astronomical sources experiences different types of transformations as it propagates through the atmosphere of the Earth and the instrument, which is shown by a schematic in Figure \ref{fig:jones_chain}. These individual effects can be expressed as a chain of Jones matrices. The overall Jones matrix, $J_\mathrm{j}$ can be written as,
\begin{equation}
\begin{aligned}
   J_\mathrm{j} (\nu,\ t,\ \vec{l}) =  G_\mathrm{j}(\nu,t)\ D_\mathrm{j}(\nu,t) E_\mathrm{jP} (\nu, t,\ \vec{l})\ F_\mathrm{j}(\nu,\ t,\ \vec{l})\\
  = G_\mathrm{j}(\nu,t)\ D_\mathrm{j}(\nu,t)\ E_\mathrm{jP,err}(\nu,\ t,\ \vec{l}) \times 
    \\E_\mathrm{jP,model} (\nu, t,\ \vec{l})\ F_\mathrm{j}(\nu,\ t,\ \vec{l})  
\end{aligned}
\label{eq1}
\end{equation}
where, index P represents the pointing direction of the telescope beam in the sky, and $\vec{l}$ represents the direction in the sky with respect to a certain reference point inside the primary beam. $G_\mathrm{j}$ and $D_\mathrm{j}$ are in the antenna frame and represent instrumental gain and leakages, respectively. $E_\mathrm{jP}$ represents the primary beam response of the telescope, which maps the electromagnetic signal from the celestial reference frame to the antenna polarization frame (X-Y / R-L) as defined by the International Astronomical Union (IAU) \citep{IAU_1973}. $E_\mathrm{jP}$ can be broken into two parts: i) $E_\mathrm{jP,model}(\vec{l})$, representing modeled or measured full Jones primary beam response and ii) $E_\mathrm{jP,err}(\vec{l})$, representing the deviation of true instrumental primary beam response from the modeled primary beam. $F_\mathrm{j}$ is the ionospheric FR, which does not affect unpolarized emission and can be ignored for an unpolarized sky. Since all these terms depend on time ($t$) and frequency ($\nu$), without loss of generality, we can rewrite the Equation \ref{eq1} for an unpolarized source at a single spectral channel and single time slice as
\begin{equation}
\begin{split}
   J_\mathrm{j} (\vec{l})= G_\mathrm{j}\ D_\mathrm{j}\ E_\mathrm{jP,err}(\vec{l})\ E_\mathrm{jP,model} (\vec{l}).\\
\end{split}
\label{eq2}
\end{equation} 

These Jones terms can be grouped into two categories -- 
\begin{enumerate}
    \item {\bf Direction dependent: }These Jones terms are function $\vec{l}$.
    \item {\bf Direction independent: }These terms are independent of $\vec{l}$. Direction-independent terms can be sub-divided into two categories -- \begin{enumerate}
        \item {\it Pointing independent: } These terms represent the effects of the backend signal chain and are independent of which direction of the sky the telescope is pointing. $G_\mathrm{j}$ and $D_\mathrm{j}$ fall in this group.
        \item {\it Pointing dependent: } These Jones terms are dependent on the pointing direction, P,  of the telescope on the sky and denoted by a subscript P.
    \end{enumerate}
\end{enumerate}
Without loss of generality, instrumental primary beam error, $E_\mathrm{jP,err}(\vec{l})$, can be factorized into a pointing-dependent component and a direction-dependent component as follows,
\begin{equation}
    E_\mathrm{jP,err}(\vec{l}) = E_\mathrm{jP,err} \Delta E_\mathrm{jP,err}(\vec{l}),
    \label{eq2a}
\end{equation}
where, $E_\mathrm{jP,err}$ representing the pointing-dependent and $\Delta E_\mathrm{jP,err}(\vec{l})$ the direction-dependent component.

Equation \ref{eq2} then can be re-written as,
\begin{equation}
\begin{split}
   J_\mathrm{j} (\vec{l})=& G_\mathrm{j}\ D_\mathrm{j}\ E_\mathrm{jP,err}\ \Delta E_\mathrm{jP,err}(\vec{l})\ E_\mathrm{jP,model} (\vec{l})\\
   =&J_\mathrm{jP}\ \Delta E_\mathrm{jP,err}(\vec{l})\ E_\mathrm{jP,model} (\vec{l})
\end{split}
\label{eq3}
\end{equation}
where all direction-independent Jones terms, both pointing dependent and pointing independent, are combined in $J_\mathrm{jP}$.

Using Equation \ref{eq3} and Equation \ref{eq:visibility_matrix}, we can write the RIME as,
\begin{equation}
\begin{split}
  \mathcal{V}_\mathrm{jk}^\prime=&J_\mathrm{j}\ \mathcal{V}_\mathrm{jk}\ J_\mathrm{k}^\dagger\\
    =&J_\mathrm{jP}\ \times \\
    &\Bigg[\int \Delta E_\mathrm{jP,err}(\vec{l})\ E_\mathrm{jP,model}(\vec{l})\ B_\mathrm{sky}(\vec{l}) \\& E_\mathrm{kP,model}(\vec{l})^\dagger\ \Delta E_\mathrm{kP,err}(\vec{l})^\dagger\ e^{\vec{b_\mathrm{jk}}. \vec{l}}\ d\vec{l}\ \Bigg] \times J_\mathrm{kP}^\dagger\\
    =&J_\mathrm{jP}\ \times \\
    &\Bigg[\int \Delta E_\mathrm{jP,err}(\vec{l})\ B_\mathrm{jk,model}(\vec{l})\ \Delta E_\mathrm{kP,err}(\vec{l})^\dagger\ e^{\vec{b_\mathrm{jk}}. \vec{l}}\ d\vec{l}\ \Bigg]\\ & \times J_\mathrm{kP}^\dagger\\
   \mathcal{V}_\mathrm{jk}^\prime=&J_\mathrm{jP}\ \mathcal{V}_\mathrm{jk,app}\ J_\mathrm{kP}^\dagger
\end{split}
\label{eq6}
\end{equation}
where, $B_\mathrm{jk,model}(\vec{l})$ represents the predicted sky model for a general array, derived using the modeled primary beams of each element. The term $\Delta E_\mathrm{jP,err}(\vec{l})$ can be neglected while computing $\mathcal{V}_\mathrm{jk,app}$ for direction-independent polarization calibration, provided that the following assumption holds, and any residual effects arising from neglecting $\Delta E_\mathrm{jP,err}(\vec{l})$ can be subsequently corrected in the image plane.
\begin{equation}
\begin{split}
    \int_l \Delta E_\mathrm{jP,err}(\vec{l})\ B_\mathrm{jk,model}(\vec{l})\ \Delta E_\mathrm{kP,err}(\vec{l})\ e^{\vec{b_\mathrm{jk}}. \vec{l}}\ d\vec{l} & \\ \approx  \int_l B_\mathrm{jk,model}(\vec{l})\ \ e^{\vec{b_\mathrm{jk}}. \vec{l}}\ d\vec{l}
\end{split}    
\label{eq:assumption}
\end{equation} 

\begin{figure*}
    \centering
    \includegraphics[trim={0.2cm 0cm 0cm 0cm},clip,scale=0.72]{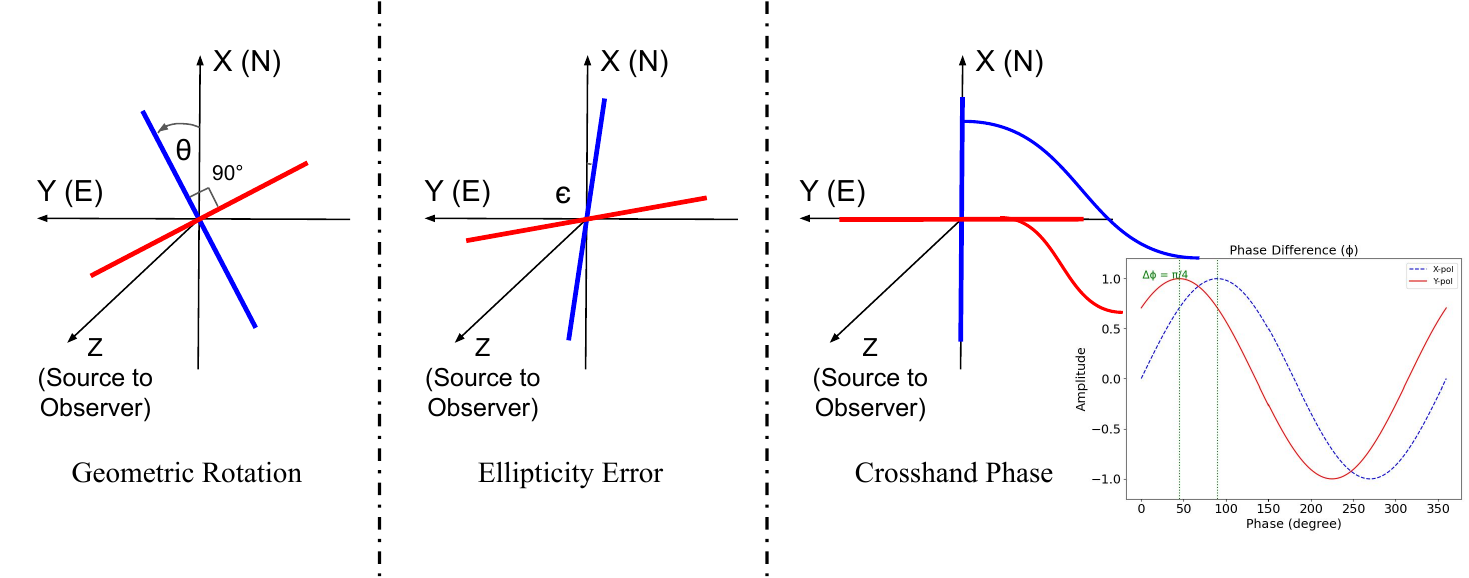}
    \caption{A schematic diagram representing origins of three pol-rotation terms in the linear polarization basis. Black arrows show the axes of the sky coordinate system as per IAU convention \citep{IAU_1973}. Bold blue and red lines represent X and Y dipoles. The first panel shows the geometrical rotation with respect to the IAU-defined antenna polarization frame by an angle $\theta$, the second panel shows the ellipticity error due to non-orthogonality by an angle $\epsilon$, and the third panel shows the phase difference by an angle $\phi$ between signals coming from X and Y dipoles.}
    \label{fig:polrot_schematic}
\end{figure*}
 
In this framework, we explicitly decompose the direction-dependent Jones terms into pointing-dependent and pointing-independent components, which are typically overlooked. However, this distinction enhances the adaptability of the formalism for both aperture arrays and dish antennas. The pointing-dependent term is particularly relevant for aperture arrays, where the instrumental primary beam response varies with the pointing direction of the telescope beam. In contrast, for dish antennas, the primary beam characteristics remain largely unchanged regardless of the pointing direction\footnote{with minor variations due to gravitational loading} and can effectively be represented by an identity matrix. 

A wide FoV instrument observing an unpolarized sky can detect multiple apparent polarized sources with distinct, linearly independent Stokes vectors. Given the high density of unpolarized sources with sufficient flux density ($>$1 Jy) at meter wavelengths, the basic requirement of at least three linearly independent Stokes vectors is easily met by the apparent sky model. Hence, the proposed method uses $\mathcal{V}_\mathrm{jk,app}$ to determine $J_\mathrm{jP}$'s. In the following subsections, we present the methodology for estimating $J_{\mathrm{jP}}$ and transferring these solutions from a field containing unpolarized sources to another field containing polarized sources.

\subsection{Choice of Appropriate Coordinate System for the Primary Beam}\label{subsec:primary_beam_coordinate}
Accurately estimating $\mathcal{V}_{\mathrm{jk,app}}$ in the antenna polarization frame from the source observed in the celestial reference frame is crucial because the $J_{\mathrm{jP}}$ matrices are computed and corrected in the antenna frame. The transformation between the celestial frame and the antenna polarization frame is encapsulated in the Jones matrix of the primary beam of the antenna, $E_{\mathrm{jP}}$.

Typically, the modeling or measurement of the primary beam of a telescope is performed in the Altitude-Azimuth (alt-az) or similar reference frame used in electromagnetic simulation software \citep[e.g.][]{Sokolowski2017}. This frame is preferred because the model primary beam remains invariant in this frame and is independent of what part of the celestial sphere is being observed.
Hence, the model beam is applicable for observations at any epoch. However, it is essential to include the necessary transformations, such as parallactic rotation, to convert from the celestial frame to the alt-az frame when applying the primary beam model for observation calibration. In Equation \ref{eq1}, $E_{\mathrm{jP}}$ must incorporate these transformations to correctly transfer the electromagnetic signal from the celestial frame to the antenna polarization frame.

\subsection{Determination of $J_\mathrm{jP}$}\label{subsec:jip_dter}
Determination of $J_\mathrm{jP}$ is generally done by minimizing the $\chi^2$ of the given form \citep{Hamaker2000},
\begin{equation}
    \chi^2=\sum_\mathrm{jk}||\mathcal{V}_\mathrm{jk}^\prime - J_\mathrm{jp}\ \mathcal{V}_\mathrm{jk,app}\ J_\mathrm{kP}^\dagger||_F
    \label{eq:RIME_minimize}
\end{equation}
where, $||.||_F$ is the Frobenious norm. It can be shown that the Frobenius norm is invariant under unitary transformations. This means Equation \ref{eq:RIME_minimize} is invariant for both $J_\mathrm{jP}$ and $J^\prime_\mathrm{jP}$ such that,
\begin{equation}
    J_\mathrm{jP} = J^\prime_\mathrm{jP}\ U_\mathrm{P}
    \label{eq:uni_invariance}
\end{equation} 
where $U_\mathrm{P}$ is a unitary matrix. 

Without loss of generality, $J_\mathrm{jP}$ can be written as the product of normalized antenna-dependent variation ($J_\mathrm{j}^\prime$) and an array averaged component, $X_\mathrm{P}$ as,
\begin{equation}
    J_\mathrm{jP}=J_\mathrm{j}^\prime\ X_\mathrm{P}
    \label{eq_norm}
\end{equation}
where, $\sum_{j}J_\mathrm{j}^\prime\approx I$. The mathematical formulation of the normalization is adapted from \cite{Kansabanik2022_paircars_alogo}. 

Any 2$\times$2 matrix always has unique polar decomposition \citep{Hamaker2000}. Hence, $X_\mathrm{P}$ can be decomposed as, $X_\mathrm{P}=xH_\mathrm{P}U_\mathrm{P}=xU_\mathrm{P}H^\prime_\mathrm{P}$, where, $x$ is a complex number, $U_\mathrm{P}$ is an unitary matrix with determinant 1, and $H_\mathrm{P}$ or $H^\prime_\mathrm{P}$ is a positive Hermitian matrix of determinant 1. The effect of $H_\mathrm{P}$ or $H^\prime_\mathrm{P}$ is the mixing between Stokes $I$ to Stokes $Q,\ U,\ V$ and hence changing the length of pol-vector without changing its orientation in pol-vector space. This Hermitian matrix is called the {\it pol-conversion} matrix. On the other hand, the unitary component causes mixing between Stokes $Q,\ U$, and $V$ without changing the length of the pol-vector. Since $\mathrm{det}(U_\mathrm{P})=1$, it belongs to $\mathbb{SU}(2)$ and is called the {\it pol-rotation} matrix \citep{Hamaker2000}. 

An important and intriguing aspect of polar decomposition is that the Hermitian component varies depending on whether it is multiplied by the unitary component from the left or the right. However, the unitary component remains unchanged \citep{Horn_Johnson_1985}. Hence, Equation \ref{eq_norm} can be re-written as,
\begin{equation}
    J_\mathrm{jP} = xJ_\mathrm{j}^\prime H_\mathrm{P}U_\mathrm{P} = xJ_\mathrm{j}^\prime U_\mathrm{P}H^\prime_\mathrm{P}
    \label{eq_norm_1}
\end{equation}
This allows one to change the position of the {\it pol-rotation} term, $U_\mathrm{P}$ in the Jones chain.

When 2$\times$2 Jones matrices are determined by minimizing Equation \ref{eq:RIME_minimize}, we effectively estimate and correct for $J^\prime_\mathrm{jP}$ given in Equation \ref{eq:uni_invariance} and a single unitary matrix, $U_\mathrm{P}$ ({\it pol-rotation} matrix) remains unconstrained. When there are at least three linearly independent Stokes visibilities available, $U_\mathrm{P}$ can be constrained if it is parameterized appropriately and suitable constraints are applied to the parameters (Appendix \ref{sec:fro_norm_uni_invar_param}). Any $2\times2$ unitary matrix in $\mathbb{SU}(2)$ (Appendix \ref{sec:fro_norm_uni_invar_param}) represents a rotation and can be decomposed into three successive rotations around three mutually perpendicular axes \citep{korn1961mathematical}. Depending on the instrumental polarization basis (linear or circular), $U_\mathrm{P}$ can be parameterized into mutually orthogonal rotations. 
\begin{itemize}
    \item {\bf Linear basis:} In the linear basis (X and Y polarization), $U_\mathrm{P} = U_V(\theta)\ U_U(\epsilon)\ U_Q(\phi)$. The decomposed terms have the form,
    \begin{equation}
       U_V(\theta) = \begin{pmatrix}
        cos\theta & sin\theta\\
        -sin\theta & cos\theta
        \end{pmatrix}
        \label{eq9}
    \end{equation} represents the geometric rotation of the feed/dipole with respect to the IAU-defined antenna polarization frame, 
    \begin{equation}
       U_U(\epsilon) = \begin{pmatrix}
        cos\epsilon & i sin\epsilon\\
        i sin\epsilon & cos\epsilon
        \end{pmatrix}
        \label{eq8}
    \end{equation} represents the ellipticity error due to non-orthogonality of dipoles, and,
    \begin{equation}
       U_Q(\phi) = \begin{pmatrix}
        e^{i\phi/2} & 0\\
        0 & e^{-i\phi/2}
        \end{pmatrix}
        \label{eq7}
    \end{equation} is the phase difference (crosshand phase) between X and Y polarization signals. The ranges of $\theta,\ \epsilon$, and $\phi$ are $(-\frac{\pi}{4}, \frac{\pi}{4}),\ (-\frac{\pi}{2}, \frac{\pi}{2})$, and $(-\pi,\pi)$. $U_V$ causes mixing between Stokes $Q$ and $U$ affecting the absolute polarization angle, $U_U$ causes mixing between Stokes $Q$ and $V$, and $U_Q$ causes mixing between Stokes $U$ and $V$. A schematic diagram is shown in Figure \ref{fig:polrot_schematic} to demonstrate these three {\it pol-rotation} terms in the linear polarization basis. 
    \item {\bf Circular basis:} In the circular basis, $U_\mathrm{v}$ transforms through the matrix, $T=\frac{1}{\sqrt{2}}\left(\begin{matrix}
        1 & i\\ 1 & -i
    \end{matrix}\right)$ as,
    \begin{equation}
        U_\mathrm{v}(\theta)=T^\dagger\ U^\mathrm{lin}_\mathrm{v}(\theta)\ T = \left(\begin{matrix}
            e^{i\theta} & 0 \\ 0 & e^{-i\theta}
        \end{matrix}\right)
    \end{equation}
    Hence, on the circular basis, Equation \ref{eq7} and Equation \ref{eq9} can be merged into a single term, $U_\mathrm{u,v}(\epsilon)$, and $U_\mathrm{P}$ can be written as\footnote{Using the cyclic property of matrix multiplication, ABC = CAB, if AC=CA}, $U_\mathrm{P}= U_\mathrm{q}(\phi)\ U_\mathrm{u,v}(\epsilon)$ and thus reduce 1 dof.
In the circular basis, the crosshand phase between two polarization channels introduces mixing between Stokes $Q$ and $U$, thereby altering the absolute polarization angle. Therefore, constraining the crosshand phase is essential for accurate calibration of the absolute polarization angle in the circular basis.
\end{itemize}

After correcting the visibilities for $J^\prime_\mathrm{jP}$ we obtain $\mathcal{V}_\mathrm{jk,cor}$. Then pol-rotations can be estimated by jointly minimizing,
\begin{equation}
    \chi^2_\mathrm{polrot} = \sum_\mathrm{jk}||\mathcal{V}_\mathrm{jk,cor}-U_\mathrm{P}\ \mathcal{V}_\mathrm{jk,app}\ U_\mathrm{P}^\dagger||_F.
\label{chi_pol_minimize}    
\end{equation}
For a well-designed instrument, the array-averaged values of $(\epsilon)$ and $(\theta)$ in the linear polarization basis are generally expected to be small and stable over time, as they arise from a systematic misalignment of the dipoles or feeds in the array relative to the sky coordinate system. These terms are also expected to be frequency-independent. In contrast, the crosshand phase $(\phi)$ can take any value and can vary both in time and frequency since it originates from delays in the signal paths of the two polarizations. For such a well-designed instrument, where only the crosshand phase is required to be constrained for absolute polarization calibration, the analytical expression derived in BY22 can be applied (a detailed derivation is provided in Appendix \ref{sec:ruby_deriviation}).

\subsection{Pointing Dependence of the Estimated Jones Terms}\label{subsec:pointing_temporal_dependence}
In Equation \ref{eq_norm_1}, the Jones terms generally depend on the pointing direction. However, for a radio interferometer comprising antenna dishes, such that the boresight of the antenna can physically be pointed towards any sky direction. The array-averaged instrumental polarization, characterized by the {\it pol-conversion} matrix, remains consistent regardless of the pointing direction. Therefore, deriving these Jones terms from one field and applying them to another will leave only the antenna-to-antenna variations uncorrected. These residual differences can be efficiently removed using polarization self-calibration (self-alignment) \citep{Hamaker2006,Kansabanik2022_paircars_alogo}.

\begin{table*}[!htbp]
    \centering
    \begin{subtable}[t]{\textwidth}
        \centering
        \begin{tabular}{|c|c|c|c|c|c|c|}
        \hline
        \textbf{Calibrator Field} & \textbf{$I$ (Jy)} & \textbf{$P$ (Jy)} & \textbf{$p$ (\%)} & \textbf{$\theta_\mathrm{source}$ (arcsec)} & \textbf{$f$ (MHz), Phase-I} & \textbf{$f$ (MHz), Phase-II Ex} \\ \hline \hline
        Cen A (core) & 9080.62 & 2.54   & 0.02  & $>$1000 &  110 & 80 \\ \hline
        Her A & 9717.09 & 1.39   & 0.01  & 1010 & 110 & 80 \\ \hline 
        Pic A & 11581.88 & 5.11  & 0.04  & 1400 & 80  & 80  \\ \hline
        3C 444 & 9564.34 & 2.89  & 0.03  & 900 & 110  & 80  \\ \hline
        Hyd A & 10410.48 & 1.94  & 0.02  & 1070 & 110  & 80 \\ \hline
        \end{tabular}
        \caption{Polarization properties and suitability of commonly used MWA calibrator fields for polarization calibration for different MWA array configurations. $I$ and $P$ represent total Stokes $I$ and linear polarization flux density within 30$^\circ$ FoV from the calibrator source. $p$ represents linear polarization fraction. $\theta_\mathrm{source}$ is the average size of the calibrator source. $f$ represents a suitable lowest frequency for MWA Phase-I and Phase-II extended configuration for calibrating all {\it pol-conversion} and {\it pol-rotation} terms.}
        \label{table:1}
    \end{subtable}
    \quad 
    \begin{subtable}[t]{\textwidth}
    \centering
    \begin{tabular}{|c|c|c|c|c|c|c|}
    \hline
    \textbf{Source Name} & \textbf{RA (hms)} & \textbf{DEC (dms)} & \textbf{RA (degree)} & \textbf{DEC (degrees)} & \textbf{$P$ (Jy)}  & \textbf{RM ($\mathrm{rad}\ m^{-2}$)}\\ \hline \hline
    GLEAM J063633-204225 & 06h 36m 33s & -20d 42m 25s & 99.141 & -20.708 & 1.10 & +50.29 \\ \hline
    GLEAM J121834-101851 & 12h 18m 34s & -10d 18m 51s & 184.642 & -10.312 & 0.71 & -8.16 \\ \hline
    GLEAM J035140-274354 & 03h 51m 40s & -27d 43m 54s & 57.942 & -27.715 & 0.52 & +33.59 \\ \hline
    GLEAM J041340+111209 & 04h 13m 40s & 11d 12m 09s & 63.448 & +11.195 & 0.39 & -13.35 \\ \hline
    GLEAM J153150+240244 & 15h 31m 50s & +24d 02m 44s & 232.960 & +24.033 & 0.37 & +11.73 \\ \hline
    \end{tabular}
    \caption{A few bright linearly polarized sources have been routinely used for estimating crosshand phase correction at the MWA. We have used two of them for verification of the estimated crosshand phase using the proposed method. These sources are part POGS-II \citep{Risley2020}. $P$ is the integrated linearly polarized flux density at 200 MHz.}
    \label{table:2}
    \end{subtable}
    \caption{Properties of calibrator fields and bright linearly polarized sources, which can be used for polarization calibration of the MWA.}
    \label{tab:main-table}
\end{table*}

For an aperture array, the above statements generally do not hold. The Jones terms in Equation \ref{eq_norm_1} can exhibit pointing dependence due to variations in the primary beam response across different sky directions. These variations, arising from discrepancies between the actual beam response and the modeled beam response ($E_\mathrm{jP,err}$) can lead to pointing-dependent behavior in both the {\it pol-conversion} and {\it pol-rotation} matrices in Equation \ref{eq3}. Consequently, it becomes challenging to transfer calibration solutions from one pointing direction to another. Among three {\it pol-rotation} terms, geometric and ellipticity errors are pointing-independent. However, it can be demonstrated (Appendix \ref{sec:beamformer_error_on_Yp}) that for an aperture array without significant beamformer delay errors among its antenna elements, which is true for the MWA \citep{Neben_2016}, the remaining {\it pol-rotation} term, crosshand phase, can also be treated as pointing-independent. In such cases, the transfer of solutions between fields primarily introduces a {\it pol-conversion} term, which results in leakage only from Stokes $I$ to Stokes $Q,\ U$, and $V$. This leakage can be mitigated by assuming that the sky at low frequencies is predominantly unpolarized, as outlined in the method described by \cite{lenc2017}.

\section{Required Conditions for the Applicability of This Formalism}\label{sec:key_conditions}
The applicability and accuracy of the proposed formalism rely on some critical factors, outlined below. Here, we have provided a first-order error evaluation.
\begin{enumerate}[label=C\arabic*., start=1]
    \item \textbf{Polarized Flux Threshold:} The predicted $\mathcal{V}_\mathrm{jk,app}$ assumes an unpolarized sky model. Any low-frequency sky is generally dominated by unpolarized sources. Due to the presence of polarized sources, the errors in the components of the $2\times2$ Jones matrices could be $\sim\frac{p}{2I}$ in the first order \citep{Sault1996_2}. Any observing field can be used for calibration, provided the error in determining gain terms causing $\frac{p}{2I}$ due to the presence of polarized sources in the field remains small, where $p$ represents the Stokes $Q,\ U$, or $V$ integrated flux across the entire field, and $I$ is the Stokes $I$ flux over the field. If one allows a tolerable error of about $2x$\% on each component of the Jones matrices, the field averaged polarization fraction in any of the Stokes $Q,\ U$, and $V$ components ($\frac{Q}{I},\ \frac{U}{I}$, or $\frac{V}{I}$) should not exceed approximately $x$\%. 
    
    \item \textbf{Linearly Independent Stokes Vectors:} At least three linearly independent Stokes vectors in $\mathcal{V}_\mathrm{jk,app}$ are necessary to constrain all 7 dof. In cases where a single bright source dominates, its beam-induced Stokes vector must vary significantly over at least three angular resolution elements to meet this requirement \citep{Sault1996_2}. However, in a special scenario when two of the {\it pol-rotation} terms, $\theta$ and $\epsilon$ are zero, the dof is reduced to 5. In that case, at-least two linearly independent Stokes vectors in $\mathcal{V}_\mathrm{jk,app}$ can be used to estimate them.
    
    \item \textbf{Beam Model Accuracy:} The accuracy of this method is closely tied to the primary beam model. In Equation \ref{eq:assumption}, we assumed that the beam-averaged error in the predicted apparent visibility, $\mathcal{V}_\mathrm{jk,app}$, is small. The integrated error in $\mathcal{V}_\mathrm{jk,app}$ due to $\Delta E_\mathrm{jP,err}$ must keep the $\frac{p}{2I}$ below $x$\% to achieve $2x$\% accuracy in all components of the Jones matrix. Thus, the accuracy of this method depends not only on the beam model but also on the presence of bright sources in regions with significant beam errors, such as sidelobes, where beam model inaccuracies are generally larger and introduce larger errors in $\mathcal{V}_\mathrm{jk,app}$.
\end{enumerate}
The error on the final Stokes visibilities depends on the errors on each of these components of the Jones matrices. The first order error on Stokes visibilities can be estimated from using Equations 36 to 43 in \cite{Sault1996_2}, and can be shown to be $x$\% if errors on all Stokes $Q,\ U$, and $V$ are assumed to be similar, $x\%$. One should carefully verify these conditions for an instrument before applying this formalism.

\section{Pol-rotation Calibration of the MWA}\label{sec:MWA_calibration}
In this section, we demonstrate the effectiveness of the proposed formalism using observations from the MWA. The results are validated by comparing them with previous methods employed at the MWA and other low-frequency aperture arrays, utilizing a linearly polarized source as detailed in \citep{Bernardi_2013}.
\begin{figure}
    \centering
    \includegraphics[width=\linewidth]{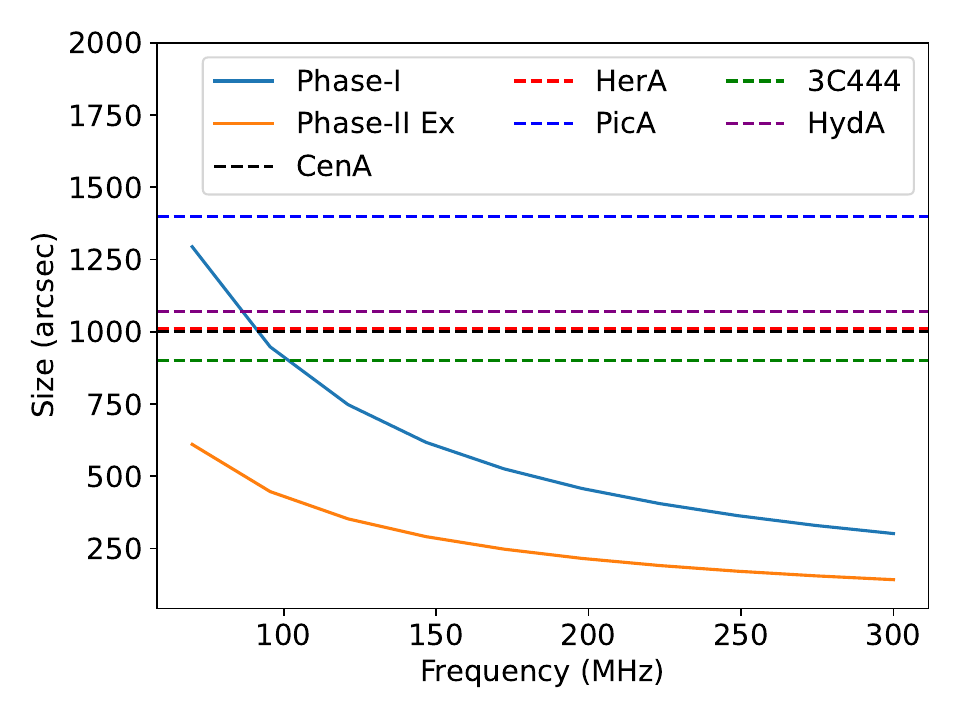}
    \caption{Average source sizes of MWA calibrator sources are shown by horizontal dotted lines. Angular size corresponding to thrice of angular resolution of the MWA Phase-I and Phase-II extended configurations are shown by blue and orange solid lines, respectively.}
    \label{fig:source_sizes}
\end{figure}

\subsection{Some Key Considerations}\label{subsec:key_consider}
The applicability and accuracy of the proposed formalism on the MWA data rely on meeting the conditions mentioned in Section \ref{sec:key_conditions}. We have evaluated these conditions for the MWA calibration below,
\begin{enumerate}
    \item \textbf{Condition C1:} Based on multiple observations with different beam pointings, we found that for most MWA calibrator fields, $\frac{p}{I} \sim2-4\%$ (Table \ref{table:1}). Thus, polarization-induced errors in Jones components are $\sim1-2\%$.  
    \item \textbf{Condition C2:} This criterion depends on spatial resolution and source size (Table \ref{table:1}, Figure \ref{fig:source_sizes}). Most MWA Phase-I calibrators (except Pic A) are $\lesssim3$ resolution elements at $\nu < 110$ MHz, limiting the determination of all three {\it pol-rotation} terms. However, assuming $\theta$ and $\epsilon$ as zero (Section \ref{subsec:geo_ellip_error}), these sources allow estimation of the remaining 5 dof (pol-conversion and crosshand phase)crosshand phase across the entire MWA observing band. 
    \item \textbf{Condition C3:} At $\nu \gtrsim 280$ MHz, grating lobes appear in the MWA tile beam \citep{Cook2021}, potentially causing errors in apparent visibilities if bright sources are present. A detailed analysis for this regime is beyond the scope of this study and will be presented in a future work. For calibrator fields at $\nu \lesssim 280$ MHz, beam-averaged residual leakages deviate from predicted models by $\sim2-4\%$.  
\end{enumerate}
Considering these factors, the suitability of the different most commonly used calibrator fields for the MWA has been evaluated and summarized in Table \ref{table:1}. These conditions must be carefully evaluated when performing polarization calibration using this method for the MWA with other observation fields.

\begin{figure*}[!htbp]
    \centering
    \includegraphics[trim={0.5cm 0.5cm 0cm 0cm},clip,scale=0.6]{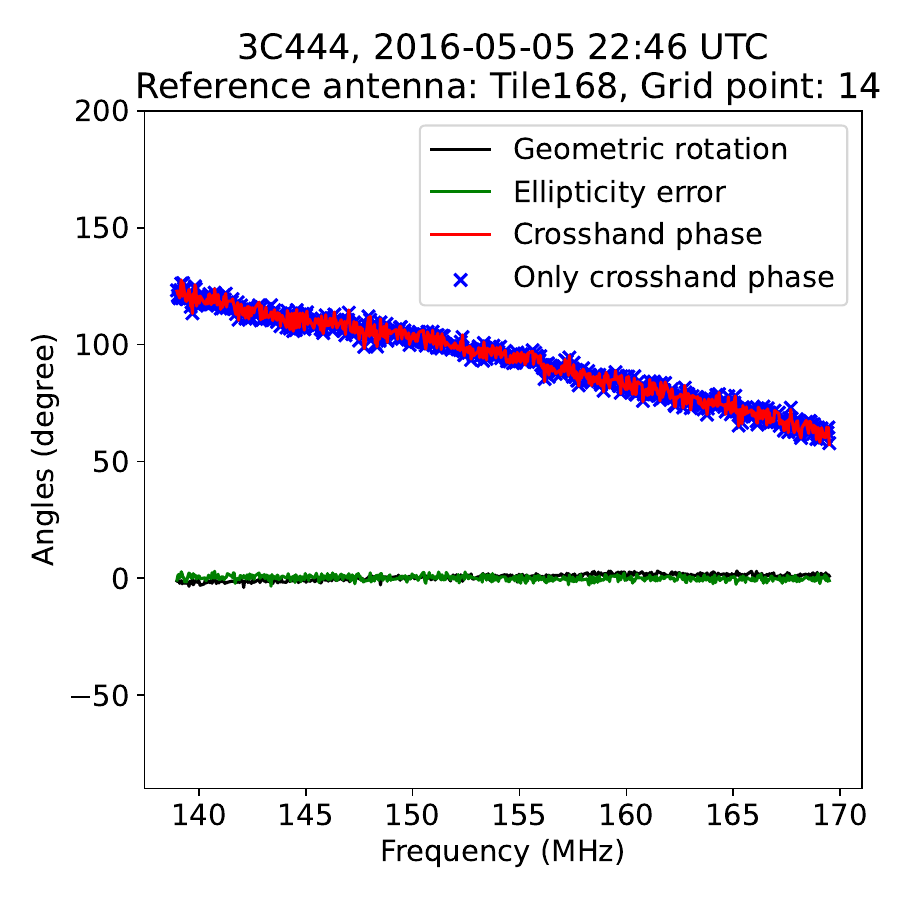}\includegraphics[trim={0.5cm 0.5cm 0cm 0cm},clip,scale=0.46]{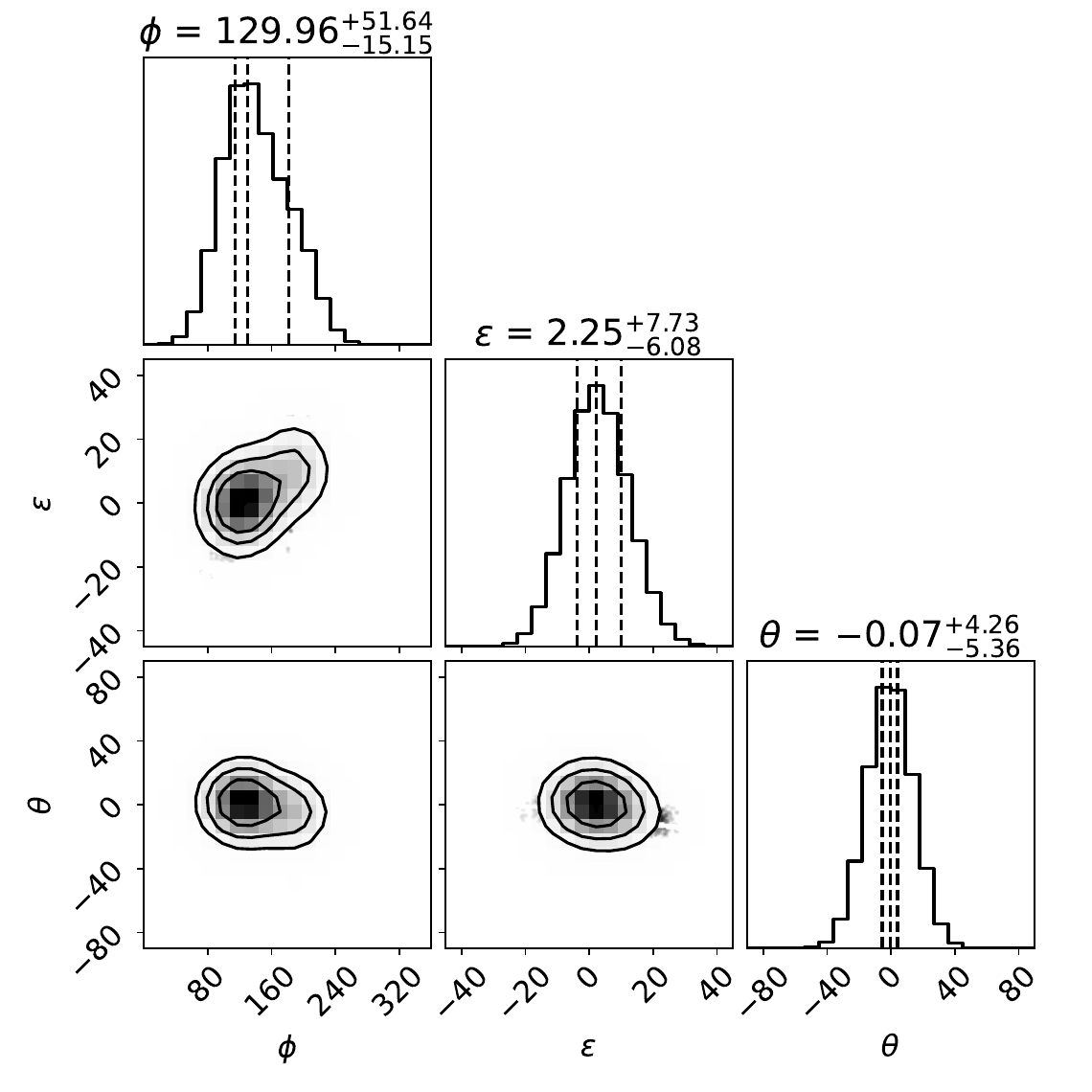}
    \caption{An example of estimated {\it pol-rotation} terms for an observation is shown. In the left panel, solid black and green lines represent the geometric angle and ellipticity errors, respectively, which remain near zero and display similar trends across the frequency range in both epochs. Blue cross markers indicate the estimated cross-hand phase assuming other {\it pol-rotation} terms are zero, closely matching the red line, which accounts for all terms. The right panel illustrates the posterior distribution of all {\it pol-rotation} terms for a single spectral channel, confirming no degeneracy among the three terms. Vertical dotted lines in the one-dimensional histograms show the 16\%, 50\%, and 84\% quantities of the distribution.}
    \label{fig:all_polrot}
\end{figure*}

\subsection{Analysis Steps}\label{subsec:analysis_steps}
The polarization calibration steps followed in this analysis are summarized below:
\begin{enumerate}
    \item Simulated $\mathcal{V}_\mathrm{jk,app}$ for the calibrator observation field following Equation \ref{eq6} using the GLEAM all-sky model \citep{Hurley-Walker2017} and the Full-Embedded Element beam model \citep{Sokolowski2017} of the MWA using \textsf{hyperdrive}\footnote{\url{https://github.com/MWATelescope/mwa_hyperdrive}}, accounting for the heterogeneous beam response due to dead dipoles in the array.
    \item Estimated 2$\times$2 Jones matrices with a chosen reference antenna by solving Equation \ref{eq:RIME_minimize}. This provides estimates of $J^\prime_\mathrm{jP}$ in Equation \ref{eq:uni_invariance}. Set off-diagonal terms of Jones matrices to zero when their real and imaginary parts were below the 1\% tolerance limit for identifying unpolarized fields. After this correction, we obtain $\mathcal{V}_\mathrm{jk,cor}$.
    \item Minimized Equation \ref{chi_pol_minimize} to estimate $\phi, \epsilon, \theta$ for individual channels after obtaining $\mathcal{V}_\mathrm{jk,cor}$. Here, we have used \textsf{scipy} minimization using a quasi-Newton minimization algorithm `L-BFGS-B' (Limited memory Broyden–Fletcher–Goldfarb–Shanno algorithm). $\epsilon$ and $\theta$ are found to be effectively zero as shown in Section \ref{subsec:geo_ellip_error}), which is expected for a well-designed instrument. For routine calibration, the analytical crosshand phase expression from BY22 (Appendix \ref{sec:ruby_deriviation}) is used to expedite analysis.
    \item Transferred 2$\times$2 Jones and crosshand phase solutions to the target field.
    \item Produced full Stokes images of the target field, corrected for the primary beam $E_\mathrm{P}(\vec{l})$ assuming a homogeneous array\footnote{Heterogeneous beam correction in the image plane is possible \citep{Cotton2021BeamCA}, which is out-of-scope of this paper}.
    \item Fitted a two-degree polynomial leakage surface to Stokes $Q,\ U$, and $V$ images following \cite{lenc2017}, assuming that most of the detected polarization arises from beam errors and residual instrumental leakage. This in principle corrects the direction-dependent beam error, $\Delta E_\mathrm{err}(\vec{l)}$.
    \item Estimated leakage surfaces for each 1.28 MHz spectral chunk are estimated for better spectral accuracy and subtracted from Stokes $Q,\ U$, and $V$ images. Spectral images within one 1.28 MHz coarse channel are corrected using the same leakage surface.
    \item Improved dynamic range by performing matrix self-alignment \citep{Hamaker2006} on the target field using the corrected $\mathcal{V}_\mathrm{jk,app}$, if necessary.
    \item Modeled and corrected ionospheric distortion using \textsf{fits\_warp} \citep{Hurley-Walker2018}, based on \cite{Hurley-Walker2017}.
\end{enumerate}
These steps are integrated into a single polarization calibration pipeline\footnote{ \href{https://github.com/devojyoti96/POLCAM/releases/tag/v1.0}{POLCAM: POLarization CAlibration routine for the MWA}}, which can be used to perform polarization calibration of MWA observations at most of the observing band ($\sim$72 -- 240 MHz) without using any bright linearly polarized source.  

\section{Results from MWA Observations}\label{sec:mwa_obs_results}
\begin{figure*}[!htbp]
    \centering
    \includegraphics[trim={0.0cm 0cm 0cm 0cm},clip,scale=0.42]{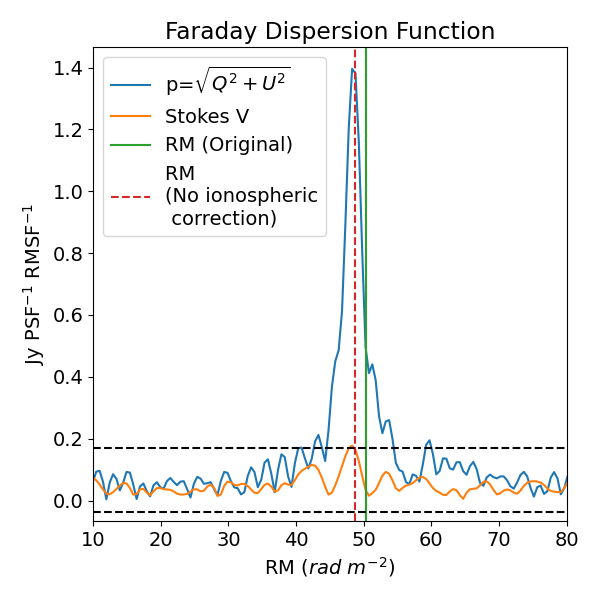}\includegraphics[trim={2.9cm 9cm 1cm 1cm},clip,scale=0.35]{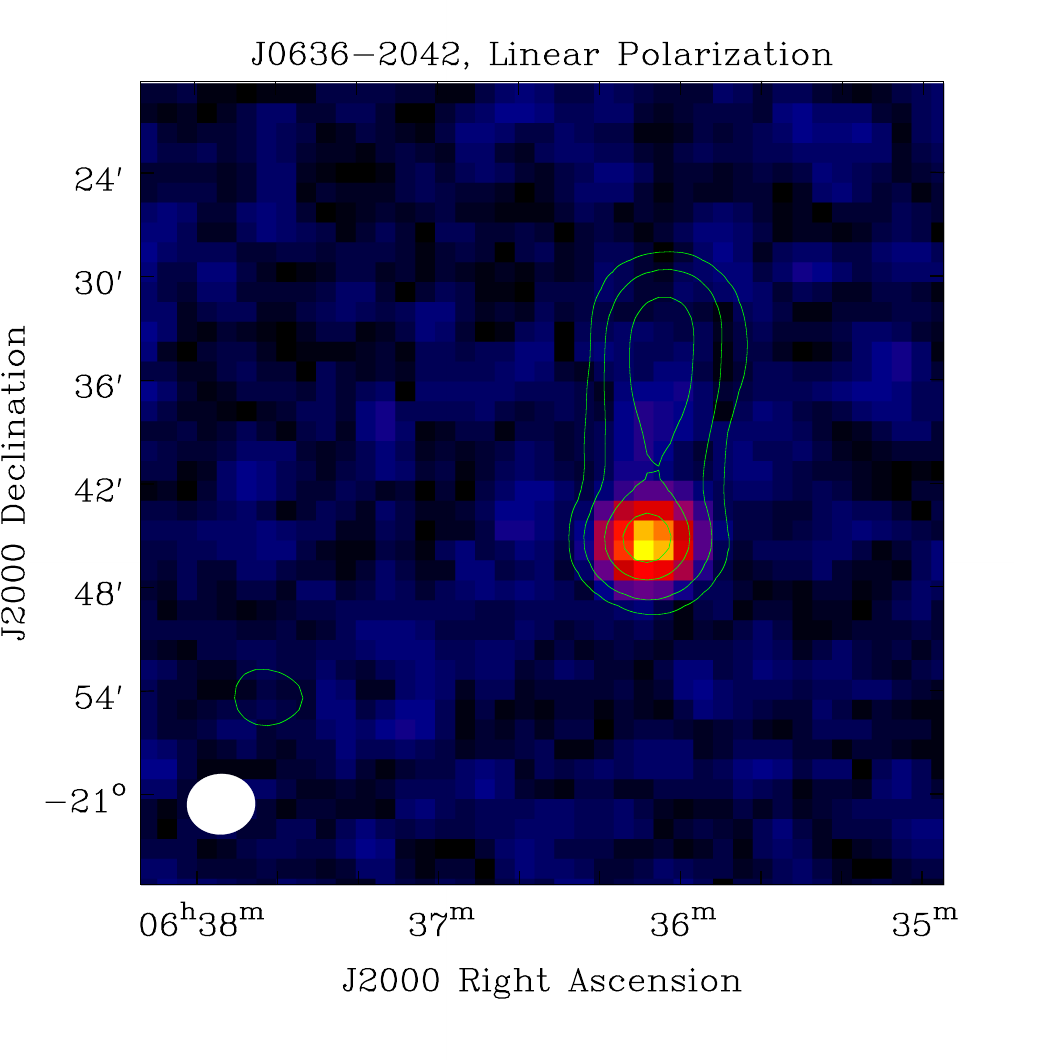}\includegraphics[trim={5.1cm 9cm 2.5cm 0cm},clip,scale=0.35]{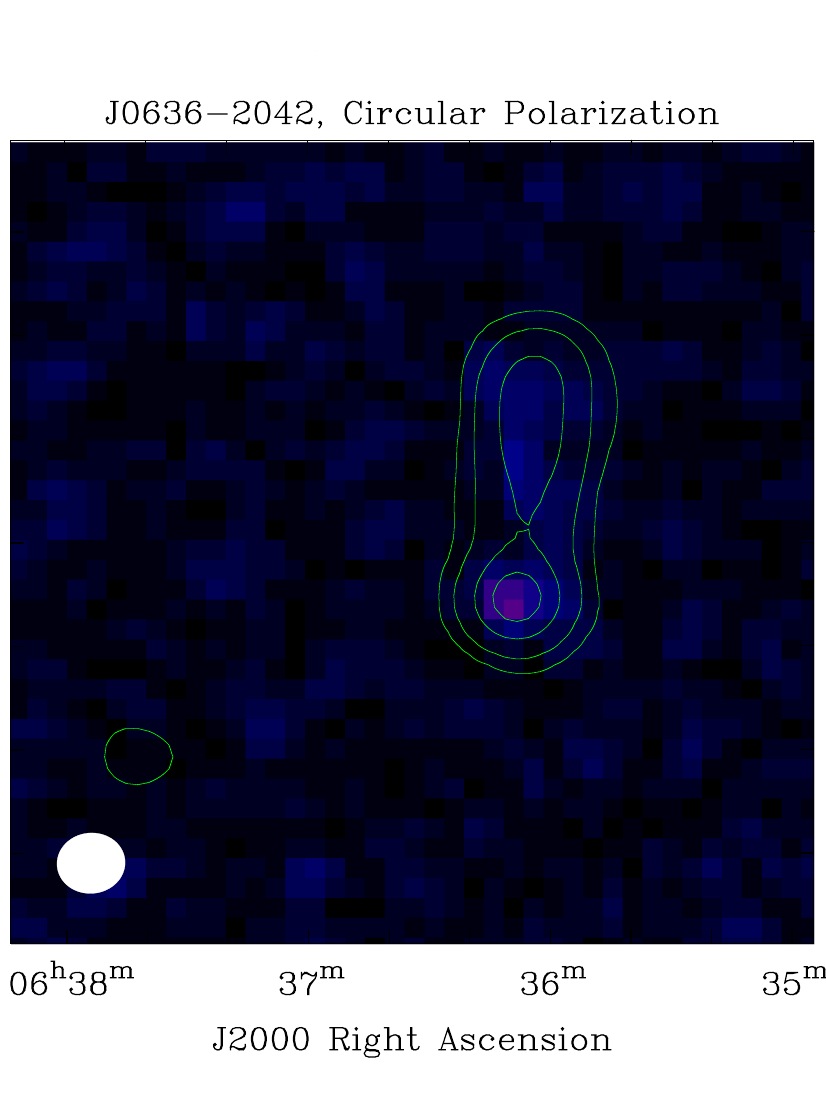}\\
    \includegraphics[trim={0.0cm 0cm 0cm 0cm},clip,scale=0.41]{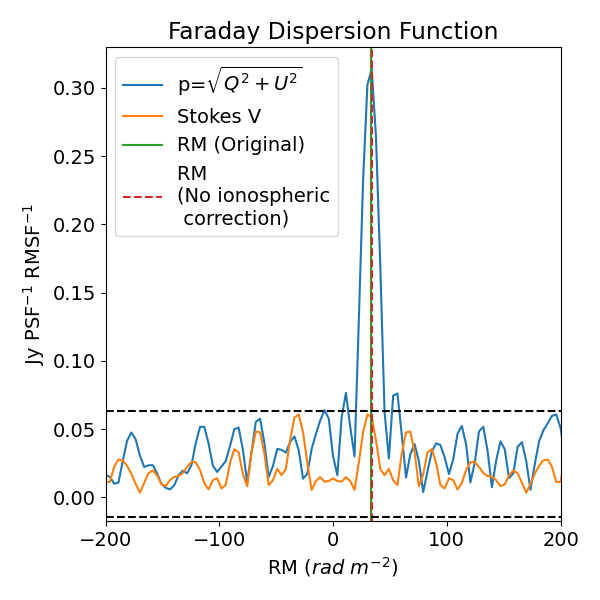}\includegraphics[trim={3cm 11.5cm 2cm 1cm},clip,scale=0.4]{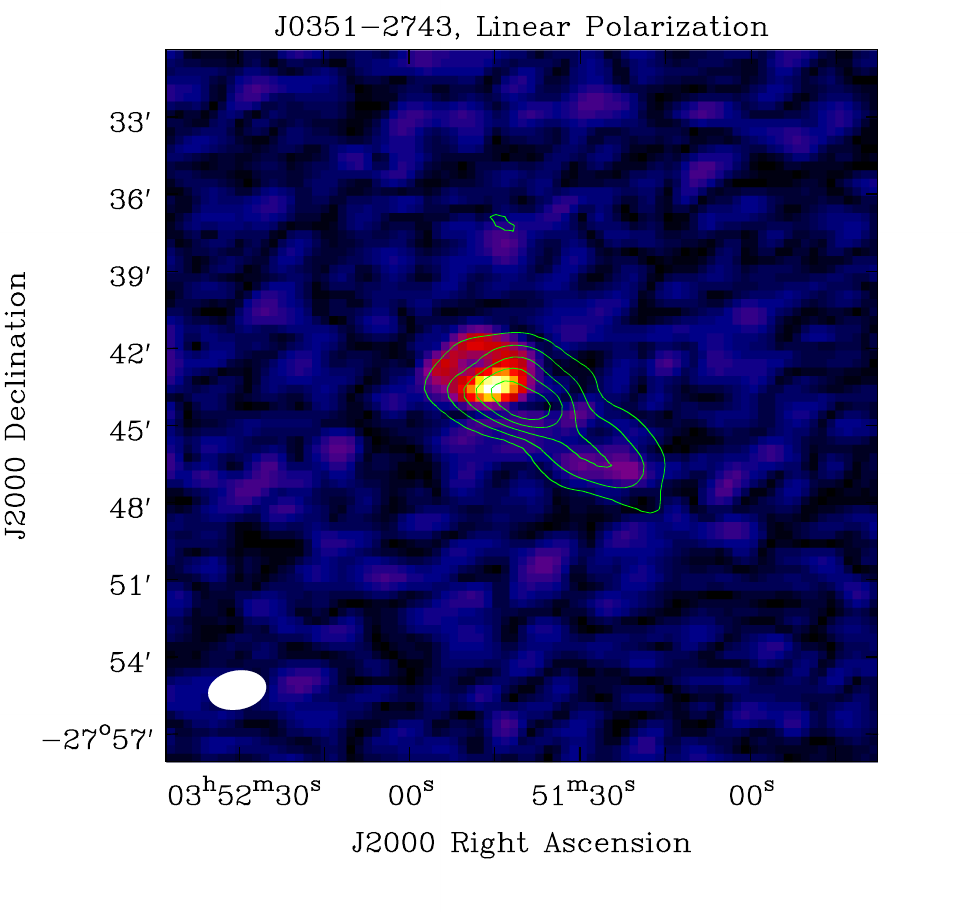}\includegraphics[trim={5.8cm 11.5cm 3cm 1cm},clip,scale=0.4]{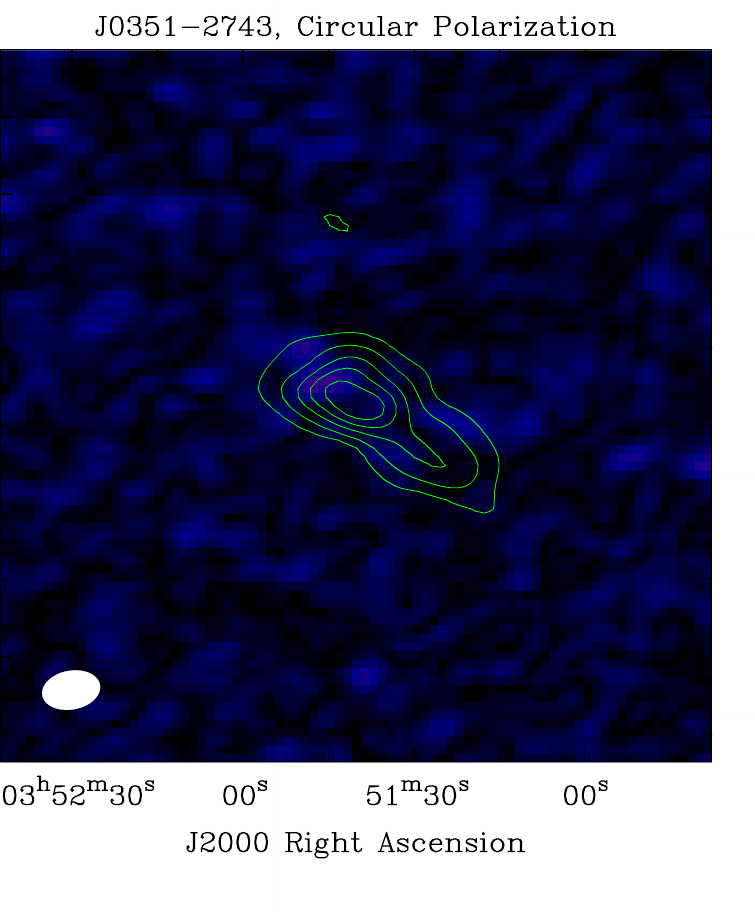}
    \caption{Validation of the new proposed method using linearly polarized sources. 1D FDFs at the peak of Stokes $I$, linear and circular polarization at the peak RM are shown respectively in three panels for GLEAM J063633-204225 (J0636-2042) in the top row and for GLEAM J035140-27435 (J0351-2743) in the middle row.  In all images, green contours are Stokes $I$ emission at 2\%, 6\%, 20\%, 40\%, 60\%, and 80\% of the peak flux density.}
    \label{fig:linpol_validate}
\end{figure*}

\begin{figure*}
\includegraphics[trim={4cm 12cm 2cm 1cm},clip,scale=0.43]{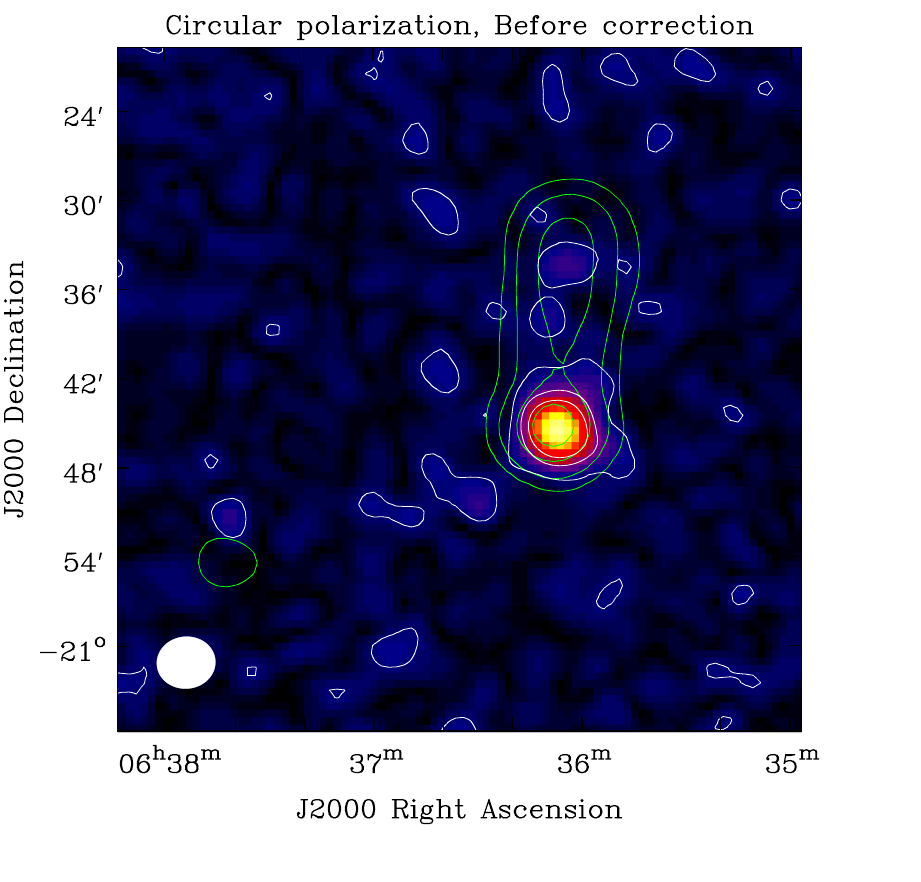}\includegraphics[trim={5.8cm 12cm 2cm 1cm},clip,scale=0.43]{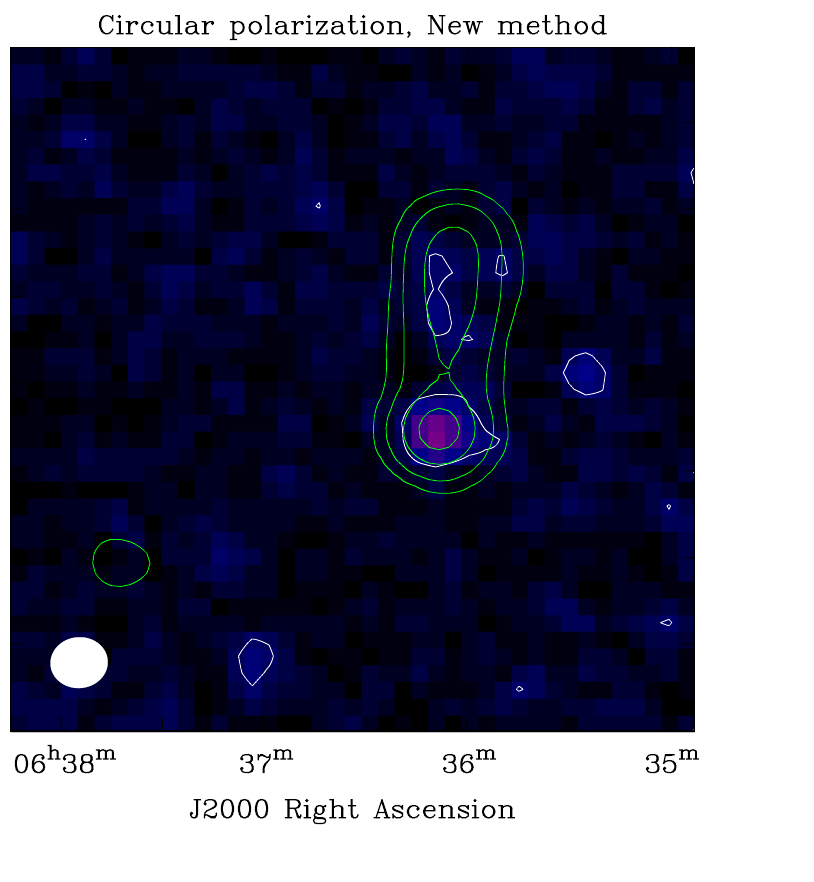}\includegraphics[trim={6cm 12cm 2cm 1cm},clip,scale=0.43]{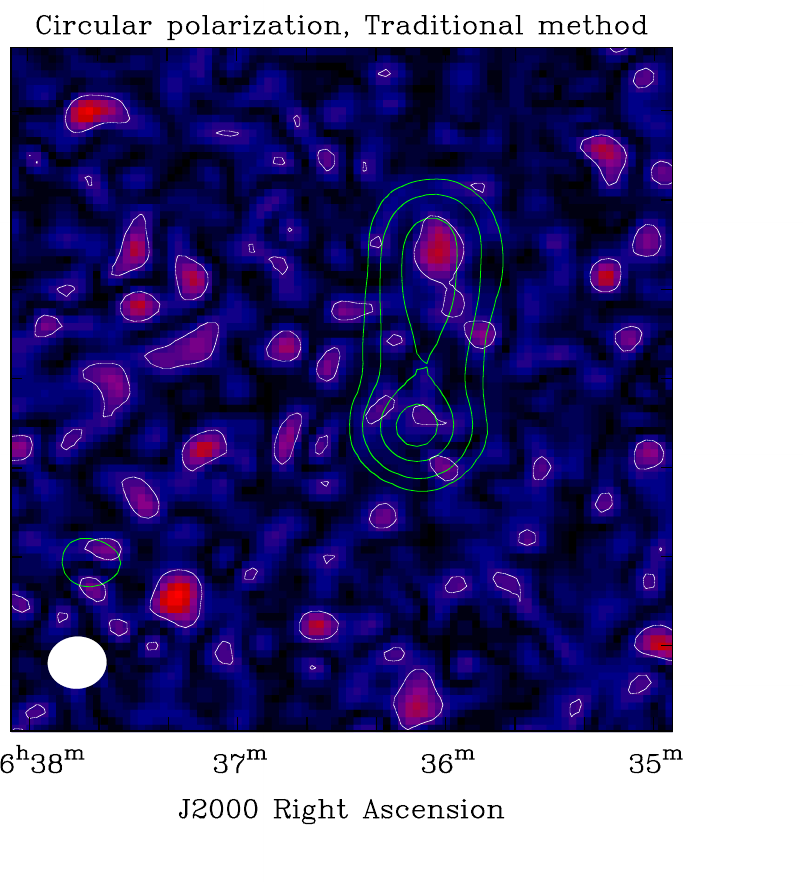}
\caption{A comparison between the traditional and newly proposed methods for crosshand phase correction is shown using J0636-2042. Three panels show Stokes $V$ images at peak RM for the three cases -- before crosshand phase correction, crosshand phase corrected using the newly proposed method, and crosshand phase corrected using the traditional method, respectively. In all images, green contours are Stokes $I$ emission at 2\%, 6\%, 20\%, 40\%, 60\%, and 80\% of the peak flux and white contours correspond to $\sigma_V,\ 3\sigma_V$, and $5\sigma_V$ of Stokes $V$ at peak RM, where $\sigma_V$ is the RMS noise level of the Stokes $V$ map at peak RM.}
\label{fig:method_comparison}
\end{figure*}
    
\subsection{Geometric Rotation and Ellipticity Errors of the MWA}\label{subsec:geo_ellip_error}
The objective of absolute polarization calibration is to align the measured visibilities to the IAU-defined antenna polarization frame and there are two geometric errors to be corrected. For a well-designed instrument like the MWA, geometric rotation, and ellipticity errors are assumed to be negligible. However, no prior observational verification of this assumption exists in the literature. As discussed in Section \ref{subsec:jip_dter}, if these two {\it pol-rotation} terms are zero, the crosshand phase can be estimated using the analytical method described in BY22. To test this, we use observation of a calibrator field of 3C444 on 2016 May 05.

Figure \ref{fig:all_polrot} presents the estimated geometric rotation (black line) and ellipticity errors (green line), which remain close to zero, independent of frequency, and consistent across epochs. The analytically estimated crosshand phases (blue crosses) align well with those derived by simultaneously solving all three {\it pol-rotation} terms (red lines). To examine potential degeneracy or local minima among the {\it pol-rotation} parameters, a Markov Chain Monte Carlo (MCMC) simulation was performed for a single spectral channel, employing 128 MCMC walkers, each with 5,000 steps, and uniform priors over the permitted parameter ranges. The resulting posterior distributions, shown in Figure \ref{fig:all_polrot}, reveal no correlations or local minima among the parameters. These findings confirm the assumptions outlined in Section \ref{subsec:jip_dter} and support the adoption of the analytical method from BY22 for the remainder of the MWA polarimetry analysis.

\subsection{Validation Using a Polarized Source}\label{subsec:valid_using_linpol}
With the geometric rotation and ellipticity error of the MWA being close to zero, the only remaining {\it pol-rotation} term to calibrate is the crosshand phase. For a polarized astronomical source with sufficient rotation measure (RM), the Faraday Dispersion Function (FDF) \citep{Brentjens2005} of linearly polarized emission should peak at the source RM, while the FDF for Stokes $V$ should not. However, if there is instrumental leakage from Stokes $U$ to Stokes $V$ due to the crosshand phase, the FDF of Stokes $V$ will show a peak at the same RM. Improper crosshand phase calibration can cause leakage between Stokes $U$ and $V$, resulting in spurious Stokes $V$ emission at the source RM. To validate the accuracy of the proposed crosshand phase calibration method, two linearly polarized sources with sufficiently large RM (from Table \ref{table:2}) are analyzed. Observation of GLEAM J063633-204225 and GLEAM J035140-274354 are at the central frequencies of 150 MHz and 223 MHz, respectively, with 30 MHz bandwidth.

After calibrating using the calibrator source field as described in Section \ref{subsec:analysis_steps} and transferring the solutions to the target source field, spectropolarimetric imaging was performed. This is followed by primary beam and {\it pol-conversion} corrections. RM-synthesis was applied separately to the Stokes $QU$ cubes and Stokes $V$ cubes. The example results for these two sources are presented in Figure \ref{fig:linpol_validate}. The left panels of the bottom and middle rows show the 1D FDFs of linear (shown by blue) and circular polarizations (shown by orange) at the peak of Stokes $I$. The green vertical solid line represents the expected RM of the source and the red vertical dotted line represents the RM value (peak-RM) of peak linearly polarized flux density. This is slightly shifted from the source RM, because no ionospheric RM correction is done. The peak linearly polarized emission is detected in 1D FDF with more than 3$\sigma$ significance (shown by black horizontal dotted lines), while circular polarization is below 3$\sigma$ significance. 

To demonstrate this is also true in the image domain, linear and circular polarization images at the peak-RM are shown in the middle and right panels, respectively. Color maps show the polarized flux density and green contours represent the Stokes $I$ emission. In the top row, linearly polarized emission is detected from the southern part of the source, while at the same location circularly polarization flux density is below 3$\sigma$ detection limit. In the bottom panel, linearly polarized emission is detected from the northern part of the source, while at the same location circularly polarization flux density is below 3$\sigma$ detection limit. The non-detection of circularly polarized emission at the peak-RM confirms the absence of residual Stokes $U$ to Stokes $V$ leakage above the noise level. This validates the accuracy of the proposed crosshand phase estimation method across multiple epochs.
\begin{figure*}[!htbp]
    \centering
    \includegraphics[width=0.95\textwidth]{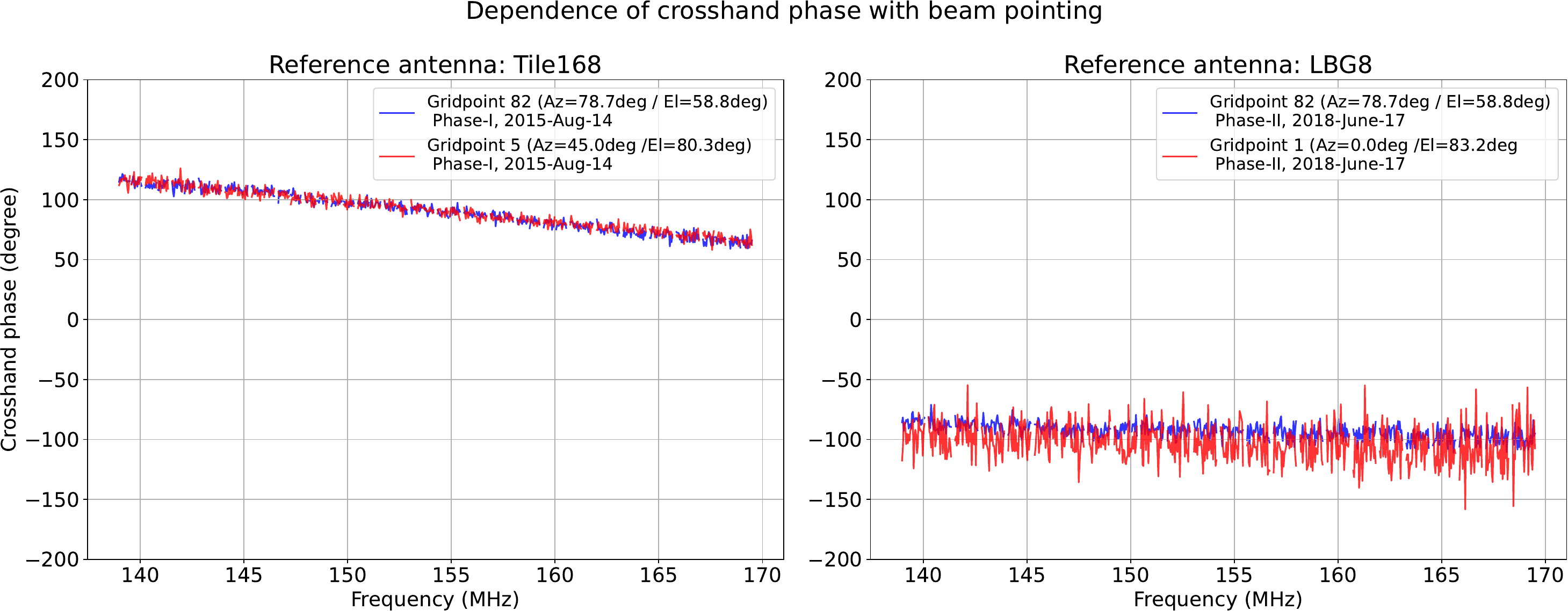}\\
   \includegraphics[width=0.95\textwidth]{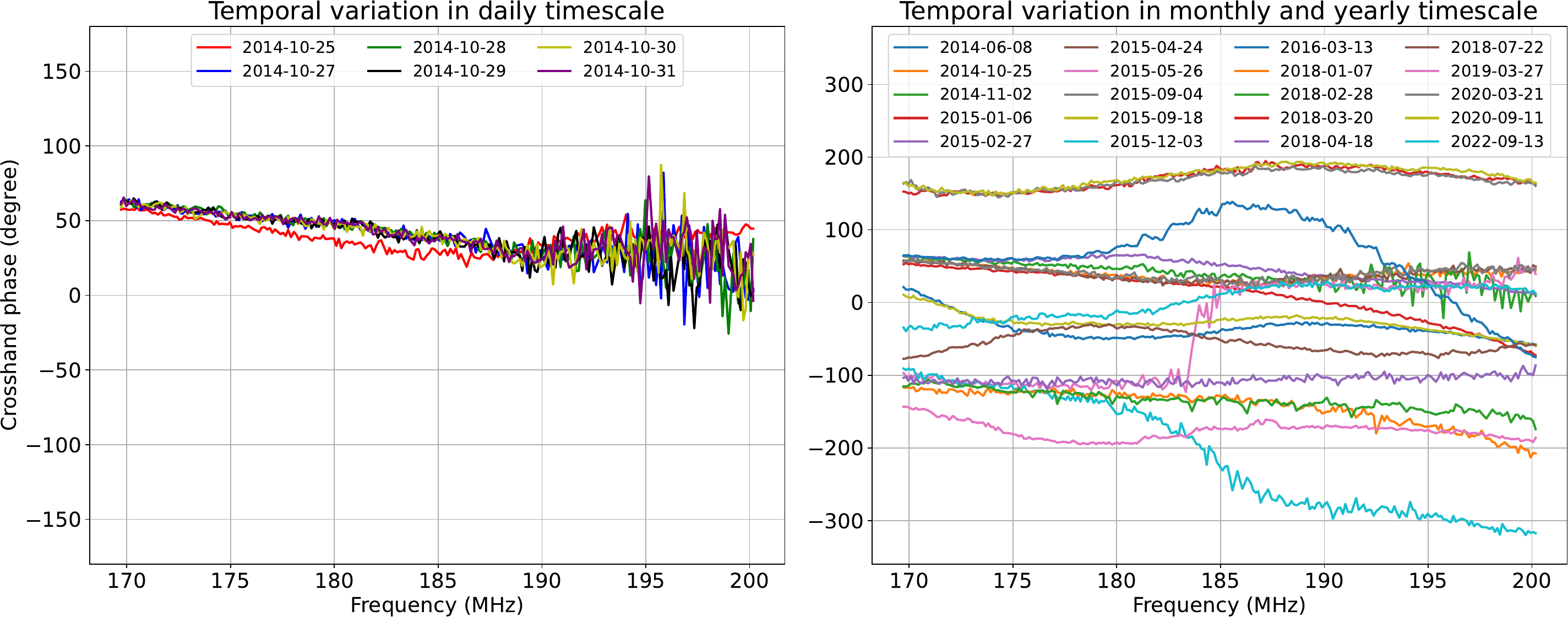}
    \caption{{\it Top row: }Crosshand phase spectra for two different beamformers pointing on the same days are shown by red and blue lines. The left panel shows observations of MWA phase-I and the right panel shows for the MWA phase-II extended configuration. Crosshand phase spectra for two different beamformer pointings at different elevations and azimuths match with each other. {\it Bottom row: }Temporal variations of MWA crosshand phase spectra from some example observations chosen randomly from the archive are shown. The same reference antenna, {\it Tile168}, has been used for all observations. The left panel shows the temporal variations on a timescale of days and the right panel shows the temporal variations on timescales spanning months to years.}
    \label{fig:pointing_temporal_dependence}
\end{figure*}

\subsection{Comparison With Traditional Method}\label{subsec:compare_trad_method}
Traditionally, a linearly polarized source is used to calibrate the crosshand phase \citep{Taylor2024}. Following primary beam correction and residual direction-dependent Stokes $I$ leakage corrections, any detected circularly polarized emission is typically attributed to leakage from Stokes $U$. Based on this assumption, the crosshand phase for each spectral channel can be estimated directly from the image using the formula:
\begin{equation}
    \phi = \mathrm{arctan}\left(\frac{V_\mathrm{flux}}{U_\mathrm{flux}}\right)
    \label{eq:crossphase_eq_linpol}
\end{equation}
where, $U_\mathrm{flux}$ and $V_\mathrm{flux}$ are Stokes $U$ and $V$ flux density. This traditional method is computationally intensive compared to the approach proposed in this work, primarily due to the requirement for spectral imaging and primary beam correction. Additionally, the method relies on adequate integration time and a sufficiently bright linearly polarized source to achieve detection at small spectral integrations.

In the traditional approach, the crosshand phase is derived in the image plane after primary beam correction and residual direction-dependent Stokes $I$ leakage correction. However, the placement of the crosshand phase Jones term differs from its position in the Jones chain employed by the newly proposed method (as shown in Equation \ref{eq_norm_1}). Consequently, the numerical values of the crosshand phases between the two methods do not align, though the final results are comparable.

We performed crosshand phase correction for GLEAM J063633-204225 using both the traditional and the proposed methods. Figure \ref{fig:method_comparison} shows the Stokes $V$ emission at the peak-RM before crosshand phase correction. In this image, Stokes $V$ emission is detected at more than $3\sigma$ significance, which is attributable to leakage from Stokes $U$ into Stokes $V$. The middle and right panels in the bottom row depict the Stokes $V$ images at the peak-RM after crosshand phase calibration using the newly proposed and traditional methods, respectively. In both cases, the Stokes $V$ images appear noise-like, validating that the proposed method achieves crosshand phase correction comparable to the traditional approach.

\subsection{Dependence of Beamformer Pointing}\label{subsec:pointing_temporal_dependence_result}
As discussed in Section \ref{subsec:pointing_temporal_dependence}, the crosshand phase estimated from a calibrator pointing can be transferred to other pointings if the crosshand phase is independent of pointing. A theoretical expectation for this independence is provided in Appendix \ref{sec:beamformer_error_on_Yp}. To validate this observationally, two sets of data from MWA phase-I and phase-II extended configurations were selected. Observations within each configuration were taken on the same day and at nearby times to minimize temporal variability. The top row of Figure \ref{fig:pointing_temporal_dependence} shows that the crosshand phase spectra for different beamformer pointings are consistent. The differences in the crosshand phase between the two configurations arise from the use of different reference antennas during calibration. These results confirm that crosshand phase spectra are indeed independent of pointing and can reliably be transferred from the calibrator to target fields. 

\subsection{Temporal Variability on Long and Short Timescales}\label{subsec:temporal_var}
To assess the stability of the crosshand phase when transferring solutions across different epochs, we analyzed calibrator fields from the MWA archive at similar spectral bands over timescales spanning the range from days to years.
Temporal variations on day timescales are shown in the bottom left panel of Figure \ref{fig:pointing_temporal_dependence}. The results indicate that the crosshand phase remains mostly stable on daily timescales of around a week. However, as shown in the bottom right panel of Figure \ref{fig:pointing_temporal_dependence}, variations are observed on monthly and yearly timescales. Though usually not, solutions separated by months may sometimes partially match for some part of the spectral range.
Similar behavior is evident on yearly timescales. These findings suggest that occasional matches in the crosshand phase across months or years are coincidental and do not imply stability over longer timescales. Further verification using similar tests on other epochs (with one example shown in the left panel) confirms that crosshand phase calibration is stable on daily timescales. This highlights the importance of performing crosshand phase calibration daily for precise polarization calibration of MWA observations, particularly when combining multiple epochs of data.

\section{Discussion}\label{sec:discussion}
Absolute polarization angle calibration is a significant challenge in low-frequency polarimetry due to the scarcity of linearly polarized calibrator sources with known polarization angles. Many calibrators used at higher frequencies \citep{Perley_2013} become depolarized at meter wavelengths. Traditionally, the community relies on either an unpolarized source along with either a linearly polarized source with a known polarization angle or a linearly polarized source whose polarization angle is not known a-priori, but is observed at least three widely separated parallactic angles \citep{evla_memo_201_hales2017} for absolute polarization angle calibration at high-frequency observations.

However, the fundamental requirement for absolute polarization calibration is the measurement of three linearly independent Stokes visibility vectors distinguishable above instrumental noise. The conventional observational approach represents one possible way to determine these three linearly independent Stokes visibility vectors and achieve absolute polarization calibration. Beam-induced polarization can also be used to perform polarization calibration for narrow FoV instruments, as demonstrated by \cite{Farnes2014} using the GMRT observations. For low radio frequency wide FoV instruments, the approach presented in this paper satisfies the fundamental requirement for enabling absolute polarization calibration. This method eliminates the need for observing a polarized source in every epoch or across a large range of parallactic angle, addressing a critical gap in low-frequency polarimetric calibration. 

The proposed approach offers an additional advantage for low-frequency polarization calibration. For linearly polarized sources, since the circular polarization is zero, the crosshand phase can be determined using Equation \ref{eq:crossphase_eq_linpol}. However, linearly polarized emission undergoes ionospheric Faraday rotation (FR), which alters Stokes $Q$ and $U$. This modified Stokes $U$ leaks into Stokes $V$. As the amount of leakage depends on the value of both the crosshand phase as well as the angle of linear polarization, estimated crosshand phase is also influenced by ionospheric FR. Although this effect is smaller at higher frequencies and is used successfully \citep{Taylor2024}, the effect of ionospheric FR becomes significant at low frequencies, making the crosshand phase direction and time dependent. 

In the proposed method, unpolarized sources are used, which are unaffected by ionospheric FR. These sources appear to be polarized only due to the polarized primary beams of the telescope, allowing the apparent polarized source model, $\mathcal{V}_\mathrm{jk,app}$, to be free of ionospheric FR. This ensures that the estimated crosshand phase is direction-independent and can be transferred across observations 
irrespective of prevailing ionospheric conditions.

We have demonstrated the applicability of this approach for MWA observations. Based on our experiments, we highlight several practical aspects to consider when calibrating the MWA observations using this method:
\begin{enumerate}
    \item We recommend using any of the calibrator fields listed in Table \ref{table:1} to estimate the crosshand phase.
    \item Other fields may also be used, provided that $\frac{p}{2I}$ is small (\(\sim 1-2\%\)). Fields containing bright linearly polarized sources often violate this condition.
    \item For the upper part of the MWA band ($\geq 280$ MHz), where the grating lobes of the primary beam are significantly strong \citep{MWA_300MHz_2021}, sources in the grating lobes should be included.
    \item Calibrator observations with integration times $<120$ seconds and close to meridian pointing do not exhibit sufficient apparent source polarization flux for estimating the crosshand phase in the presence of instrumental noise. 
    \item Suitable large spectral integration or a three-degree polynomial fitting on the crosshand spectrum can be used to obtain a smooth spectral response of crosshand phase.
\end{enumerate}

\section{Conclusion}\label{sec:conclusion}
Low-frequency polarization observations have significant potential to address key scientific questions in astronomy and astrophysics \citep[e.g.][]{lenc2017}, as well as in space weather research \citep[e.g.][]{oberoi2012,vourlidas2020,Kansabanik2024_heliopolarimetry} and terrestrial ionospheric studies. However, the scarcity of polarized sources limits the applicability of conventional high-frequency polarization calibration methods \citep{Hales_2017} at low frequencies. The concept of using apparent polarization from unpolarized sources was initially proposed by \cite{Farnes2014} and \cite{Byrne2022} for specific telescopes. Our study establishes a mathematical framework, grounded in the first principles of radio polarimetry \citep{Hamaker1996_1,Hamaker2000}, to generalize and extend these approaches to any low-frequency radio interferometers such as MWA, LOFAR, NenuFAR, OVRO-LWA, uGMRT and Square Kilometre Array Observatory (SKAO). The success of this method relies critically on accurate knowledge of the instrumental primary beam. In summary, absolute polarization calibration can be achieved by obtaining calibration information from either astronomical sources or precise instrumental models.

\cite{Byrne2022} highlighted the importance of validating polarization calibration using the apparent polarization of unpolarized sources against bright, linearly polarized sources. In this work, we utilized two bright linearly polarized sources to demonstrate the accuracy and applicability of the proposed method for estimating crosshand phases in MWA observations. Beyond crosshand phase estimation, this study extends the method to determine all three {\it pol-rotation} terms required for absolute polarization calibration. For well-designed instruments, array-averaged geometric and ellipticity errors are generally assumed to be zero. This work provides the first observational verification of these assumptions for any low-frequency radio interferometer, confirming that the absolute polarization angle aligns with the telescope reference frame and the sky coordinate system. We anticipate extending this approach to other low-frequency radio telescopes in the future.

\appendix
\section{Parameterized Minimization of the Frobenius Norm}\label{sec:fro_norm_uni_invar_param}
Since the Frobenius norm is unitary invariant, minimizing Equation \ref{chi_pol_minimize} does not generally yield a unique $U$ unless it is properly parameterized. This non-uniqueness arises due to several factors, including global phase ambiguity and symmetries in $\mathcal{V}_\mathrm{jk,cor}$ and $\mathcal{V}_\mathrm{jk,app}$. To address these issues, $U$ is parameterized such that $U \in \mathbb{SU}(2)$, where $\mathbb{SU}(2)=\{U \in \mathbb{C}^{2\times2} \mid U^*U=I, \ \mathrm{det}(U)=1\}$.

\subsection{Global phase ambiguity}\label{subsec:global_phase_ambiguity} 
Any unitary matrix, $U$ belongs to $\mathbb{U}(2)$ group can be written as, $UU^\dagger=I$. The determinant of $U$ is a complex number of modulus 1, $\mathrm{det}(U) \in \mathbb{U}(1)$, representing a global phase factor $e^{i\theta}$. Hence, any $U\in\mathbb{SU}(2)$ can be written as,
\begin{equation}
    U = e^{i\theta} V,
\end{equation}
where, $V\in \mathbb{SU}(2)$, i.e., $\mathrm{det}(V)=1$. 

If one put the constraint, $U\in\mathbb{SU}(2)$, 
\begin{equation}
    \mathrm{det}(U)=e^{i\theta}\mathrm{det}(V) = e^{i\theta} 
\end{equation}
The condition, $U\in\mathbb{SU}(2)$ makes $\mathrm{det}(U)=1$ and restricts $\theta$ to be 0 and removing global phase ambiguity.

\subsection{Symmetry breaking}\label{subsec:symmetry_breaking} 
Assume a matrix, $X$, that exhibits any particular symmetry, such as invariance under certain transformations or possessing a block diagonal, real, or Hermitian form. This implies, $X$ commute with the certain group of transformations, $G$, such that, 
\begin{equation}
gXg^{\dagger}=X, 
\label{eq:a9}
\end{equation}
where, $g\in G$. Let us now consider a unitary transformation of the form, 
\begin{equation}
Y=UXU^\dagger,
\label{eq:a10}
\end{equation}
which relates $Y$ and $X$, where, $U$ is a unitary matrix. We will prove that, by requiring $U \in \mathbb{SU}(2)$, the symmetries by $G$ are broken through asymmetry introduced by $\mathbb{SU}(2)$, because $U$ does not necessarily commute with $g\in G$. After the transformation of $X$ to $Y$, to maintain symmetry of $Y$ under the same group $G$ would require,
\begin{equation}
    gYg^\dagger= Y 
    \label{eq:a11}
\end{equation}

Expanding the left hand side of Equation \ref{eq:a11},
\begin{equation}
gYg^\dagger = g(UXU^\dagger)g^\dagger = (gU)X(Ug)^\dagger
\label{eq:a12}
\end{equation}
Substituting Equation \ref{eq:a12} and \ref{eq:a10} in Equation \ref{eq:a11} and using the Unitary property, $UU^\dagger=U^\dagger U = I$,
\begin{equation}
\begin{split}
  (gU)X(gU)^\dagger &= UXU^\dagger
\end{split}
\label{eq:a13}
\end{equation}
Now, substituting symmetry property of $X$ from Equation \ref{eq:a9},
\begin{equation}
\begin{split}
    (gU)X(gU)^\dagger &= U(gXg^\dagger) U^\dagger\\
    (gU)X(gU)^\dagger &= (Ug) X (Ug)^\dagger
\end{split}
\label{eq:a14}
\end{equation}
This shows that to maintain the same symmetry under group $G$ for $Y$ as shown in Equation \ref{eq:a11}, the equality in Equation \ref{eq:a14} must hold. This equality holds when $gU=Ug$. In general, $gU\neq Ug$, hence $Y$ does not necessarily maintain the same symmetry under group $G$ as $X$. This proves that the Unitary transformation on $X$ breaks the symmetry. 

Since the Euler basis is a complete basis representing $\mathbb{SU}(2)$ group, parameterizing $U$ in any Euler angle representation within $\mathbb{SU}(2)$ breaks symmetries in $X$ and $Y$. The parameterization in Equations \ref{eq7}, \ref{eq8}, and \ref{eq9} represents a special form of the Euler angle representation. Specifically:
\begin{itemize}
    \item $U_\mathrm{Q}(\phi)$ applies a relative phase shift between the polarization basis vectors, breaking symmetries tied to equal phases in $X$.
    \item $U_\mathrm{V}(\theta)$ rotates the eigenbasis of $X$, mixing diagonal and off-diagonal components, breaking diagonal symmetry.
    \item $U_\mathrm{U}(\epsilon)$ introduces an additional rotation around the eigenbasis, breaking any remaining symmetry.
\end{itemize}

Thus, parameterizing $U$ within $\mathbb{SU}(2)$ ensures symmetry breaking and resolves the invariance issue. However, minimizing Equation \ref{chi_pol_minimize} with respect to $\phi, \theta, \epsilon$ does not guarantee unique real parameter solutions, since $\phi, \phi+2m\pi$; $\theta, \theta+(2n+1)\pi$; and $\epsilon, \epsilon+2m\pi$ are all valid degenrate solutions, where $m\neq 0$ and $n>0$ are integers. By restricting angular ranges as described in Section \ref{subsec:jip_dter}, and combining this with the parameterization, the solutions for $\phi, \epsilon, \theta$ become unique.

\section{Dependence of Crosshand Phase on Beamformer Errors}\label{sec:beamformer_error_on_Yp}
Let us assume the electric field pattern of a single dipole and single polarization is given by,
\begin{equation}
    E_i(t)=E_ie^{j(\omega_0t+\phi_i)}
    \label{eq:electric_field}
\end{equation}
where $E_i$ is the amplitude, $\omega_0$ is the angular frequency, and $\phi_i$ is an additional phase. The beamformed signal can then be expressed as,
\begin{equation}
    E_i^{\mathrm{beam}}(t)=\sum_i w_i E_i(t-\tau_i)
    \label{eq:beamformer_signal}
\end{equation}
where $w_i$ is the beamformer weight, and $\tau_i$ is the time delay applied to align signals from different antenna elements. Substituting Equation \ref{eq:electric_field} into Equation \ref{eq:beamformer_signal} for both polarizations, we get:
\begin{equation}
    E_{X}^{\mathrm{beam}}(t)=\sum_i w_i^X E_i^Xe^{j[\omega_0(t-\tau_i)+\phi_i^X]}
    \label{eq:polx_beamformer}
\end{equation}
\begin{equation}
    E_{Y}^{\mathrm{beam}}(t)=\sum_i w_i^Y E_i^Ye^{j[\omega_0(t-\tau_i)+\phi_i^Y]}
    \label{eq:poly_beamformer}
\end{equation}
Error in $E_p^{\mathrm{beam}}$ may arise due to inaccuracies in modeling $w_i$ and $\tau_i$:
\begin{enumerate}
    \item $w_i$ are simulated weights for individual antenna elements, determined by their positions and mutual coupling in the beamformer. Since $w_i$ is pointing-independent, the error introduced by $w_i$ is also pointing-independent. However, different errors in $w_i$ for X and Y polarizations can introduce an additional phase difference between the two polarizations, which is pointing-independent.
    \item To point toward different parts of the sky, different combinations of $\tau_i$ are required. Thus, the errors in $\tau_i$ propagate differently depending on the location in the sky.
\end{enumerate}
For the MWA, these two errors have distinct effects:
\begin{enumerate}
    \item When a dipole of one polarization fails while the other polarization functions, the weights for the two beamformer signals differ. If that tile is used as a reference tile, this will introduce a frequency-dependent crosshand phase that is pointing-independent but varies from epoch to epoch due to faulty dipoles. 
    \item For MWA, \cite{Neben_2016} measured an rms beamformer group delay error per polarization of 54 ps and 21 ps for the maximum and minimum delays of the beamformer, respectively. These delays correspond to the lowest elevation and zenith pointing, respectively. Assuming independent rms errors for both polarizations, the maximum array averaged delay error (128 tiles) is approximately $\sim 0.8$ ps, leading to an array-averaged XY-phase error of $\sim 0.04$, 0.08, and 0.14 degrees at 80, 150, and 300 MHz, respectively.
\end{enumerate}
This calculation demonstrates that, for the MWA, the array-averaged crosshand phase may vary from epoch to epoch due to faulty dipoles in a significant number of tiles. However, the pointing-dependent error remains small ($\lesssim 1$ degree).

\section{Derivation of Analytical Expression of Crosshand Phase}\label{sec:ruby_deriviation}
BY22 provided an analytical expression to estimate the crosshand phase when other {\it pol-rotation} terms are neglected. Here, we provide a detailed derivation of that analytical expression starting from the minimization equation. Without loss of generality, Equation \ref{eq:RIME_minimize} can be rewritten to include all four correlation products of an interferometer. Considering only diagonal gain terms and the crosshand phase while ignoring other {\it pol-rotation} and {\it pol-conversion} terms, Equation \ref{eq:RIME_minimize} becomes:
\begin{equation}
\begin{split}
\chi^2 &= \sum_\mathrm{jk}\sum_{ab} w_\mathrm{jk}\big| \mathcal{V}_{\mathrm{jkab}}^\prime - g_{\mathrm{ja}} g_{\mathrm{kb}}^* \mathcal{V}_{\mathrm{jkab,app}} \big|^2 \\
&= \sum_\mathrm{jk} w_\mathrm{jk} \Bigg[ 
\big| \mathcal{V}_{\mathrm{jkaa}}^\prime - g_{\mathrm{ja}} g_{\mathrm{kb}}^* \mathcal{V}_{\mathrm{jkbb,app}} \big|^2 + \\
&\quad \big| \mathcal{V}_{\mathrm{jkaa}}^\prime - g_{\mathrm{jb}} g_{\mathrm{kb}}^* \mathcal{V}_{\mathrm{jkbb,app}} \big|^2 + \\
&\quad \big| \mathcal{V}_{\mathrm{jkab}}^\prime - g_{\mathrm{ja}}^\prime e^{i\frac{\phi}{2}} g_{\mathrm{kb}}^{\prime *} e^{i\frac{\phi}{2}} \mathcal{V}_{\mathrm{jkab,app}} \big|^2 + \\
&\quad \big| \mathcal{V}_{\mathrm{jkba}}^\prime - g_{\mathrm{jb}}^\prime e^{-i\frac{\phi}{2}} g_{\mathrm{ka}}^{\prime *} e^{-i\frac{\phi}{2}} \mathcal{V}_{\mathrm{jkba,app}} \big|^2
\Bigg],
\end{split}
\label{eq:RIME_minimize_allpol}
\end{equation}
where $a,\ b$ are the two polarizations ($X, Y$ or $R, L$), $w_\mathrm{jk}$ is a scalar weight for each visibility, and $g_a=g^\prime_a e^{i\frac{\phi}{2}},\ g_b=g^\prime_b e^{-i\frac{\phi}{2}}$. Minimizing Equation \ref{eq:RIME_minimize_allpol} with respect to $\phi$ reduces it to Equation 30 of BY22. In BY22, $w_\mathrm{jk}=1$ is assumed.
\begin{equation}
\begin{aligned}
    \chi^2=& \sum_\mathrm{jk} w_\mathrm{jk} \left [ \left |\mathcal{V}_\mathrm{jkab}-e^{-i \phi}M_\mathrm{jkab} \right |^2 + \left |\mathcal{V}_\mathrm{jkba}-e^{i \phi}M_\mathrm{jkba} \right |^2 \right ]\\
    = & \Delta_1+\Delta_2,
\end{aligned}
\label{eq:main_eq}
\end{equation}
where $M_\mathrm{jkab} = {g^\prime}_{ja}{g^\prime}_{kb}^*\mathcal{V}_{\mathrm{jkab,app}}$ and $M_\mathrm{jkba} = {g^\prime}_{jb}{g^\prime}_{ka}^*\mathcal{V}_{\mathrm{jkba,app}}$ are visibilities obtained after corrupting the apparent model visibilities with the estimated complex gains. 

Expanding $\Delta_1$ and $\Delta_2$, we have:
\begin{equation}
\begin{split}
    \Delta_1=& \sum_\mathrm{jk} w_\mathrm{jk} \left |\mathcal{V}_\mathrm{jkab}-e^{-i \phi}M_\mathrm{jkab} \right |^2\\
    =& \sum_\mathrm{jk} w_\mathrm{jk} \left (\mathcal{V}_\mathrm{jkab}-e^{-i \phi}M_\mathrm{jkab} \right )\left (\mathcal{V}_\mathrm{jkab}^{*}-e^{i \phi}M_\mathrm{jkab}^{*} \right )
\end{split}
\label{eq:phi1}
\end{equation}

\begin{equation}
\begin{split}
\Delta_2=& \sum_\mathrm{jk} w_\mathrm{jk} \left |\mathcal{V}_\mathrm{jkab}-e^{i \phi}M_\mathrm{jkba} \right |^2\\
=& \sum_\mathrm{jk} w_\mathrm{jk} \left (\mathcal{V}_\mathrm{jkba}-e^{i \phi}M_\mathrm{jkba} \right )\left (\mathcal{V}_\mathrm{jkba}^*-e^{-i \phi}M_\mathrm{jkba}^* \right ).
\end{split}
\label{eq:phi2}
\end{equation}

Evaluating the derivatives of $\Delta_1$ and $\Delta_2$ with respect to $\phi$, we obtain:
\begin{equation}
\begin{split}
    \frac{\delta \Delta_1}{\delta \phi}=& - w_\mathrm{jk} 2i \bigg[\mathrm{Re}(\mathcal{V}_\mathrm{jkab}M_\mathrm{jkab}^*)\sin \phi \\& + \mathrm{Im}(\mathcal{V}_\mathrm{jkab}M_\mathrm{jkab}^*)\cos \phi \bigg].
\end{split}
\label{eq:delphi1_del_delta}
\end{equation}

\begin{equation}
\begin{split}
    \frac{\delta \Delta_2}{\delta \phi}=& w_\mathrm{jk} 2i \bigg[-\mathrm{Re}(\mathcal{V}_\mathrm{jkba}M_\mathrm{jkba}^*)\sin \phi \\&+ \mathrm{Im}(\mathcal{V}_\mathrm{jkba}M_\mathrm{jkba}^*)\cos \phi \bigg].
\end{split}
\label{eq:delphi2_del_delta}
\end{equation}

The crosshand phase $\phi$, which minimizes Equation \ref{eq:main_eq}, satisfies:
\begin{equation}
\begin{aligned}
    \tan \phi = \frac{\sum_\mathrm{jk} \mathrm{Im}[w_\mathrm{jk}(\mathcal{V}_\mathrm{jkab}^*M_\mathrm{jkab} + \mathcal{V}_\mathrm{jkba}M_\mathrm{jkba}^*)]}{\sum_\mathrm{jk} \mathrm{Re}[w_\mathrm{jk}(\mathcal{V}_\mathrm{jkab}^*M_\mathrm{jkab} + \mathcal{V}_\mathrm{jkba}M_\mathrm{jkba}^*)]}.
\end{aligned}
\end{equation}

Finally, the analytical expression for the crosshand phase $\phi$ is:
\begin{equation}
    \phi=\mathrm{Arg} \Bigg [\sum_\mathrm{jk}\left[w_\mathrm{jk}(\mathcal{V}_\mathrm{jkab}^*M_\mathrm{jkab}+\mathcal{V}_\mathrm{jkba}M_\mathrm{jkba}^*)\right]\Bigg ].
    \label{eq:final}
\end{equation}
This is the same as Equation 31 of BY22 when all $w_\mathrm{jk}$ = 1.

\begin{acknowledgements}
D. K. acknowledges support for this research by the NASA Living with a Star Jack Eddy Postdoctoral Fellowship Program, administered by UCAR’s Cooperative Programs for the Advancement of Earth System Science (CPAESS) under award 80NSSC22M0097. A. V. is supported by NASA grant \#80NSSC22K1028. S. D. and D. O. acknowledge the support of the Department of Atomic Energy, Government of India, under project no. 12-R\&D-TFR-5.02-0700.  D. K. acknowledges the insightful discussions with Xiang Zhang and John Morgan. This scientific work uses the Murchison Radio-astronomy Observatory (MRO), operated by the Commonwealth Scientific and Industrial Research Organisation (CSIRO). We acknowledge the Wajarri Yamatji people as the traditional owners of the Observatory site. Support for the operation of the MWA is provided by the Australian Government's National Collaborative Research Infrastructure Strategy (NCRIS), under a contract to Curtin University administered by Astronomy Australia Limited. We acknowledge the Pawsey Supercomputing Centre, which is supported by the Western Australian and Australian Governments. We thank the reviewer for the careful review of the work which improves the quality of the presentation.
\end{acknowledgements}

\bibliography{sample631}{}
\bibliographystyle{aasjournal}

\end{document}